%% file: UN271.tex
\documentclass[]{aastex7}
\usepackage{ulem}

\newcommand\subs[1]{\textsubscript{#1}}
\newcommand\sups[1]{\textsuperscript{#1}}
\newcommand\rh[1]{\textcolor{black}{{\textit{r}\subs{H}}#1}}
\newcommand\lp[1]{\textcolor{black}{{\textit{L}\subs{p}}#1}}

\newcommand\Ju[1]{\textcolor{black}{{\textit{J}}#1}}
\newcommand\ie[1]{\textcolor{black}{{\textit{i.e.,}}#1}}
\newcommand\eg[1]{\textcolor{black}{{\textit{e.g.,}}#1}}

\newcommand\kms[1]{\textcolor{black}{{km\,s$^{-1}$}#1}}

\definecolor{gold}{rgb}{0.64,0.54,0.29}

\received{2025 April 16}
\revised{2025 May 2}
\accepted{2025 May 7}
\submitjournal{The Astrophysical Journal Letters}

\shorttitle{First Detection of Molecular Activity in C/2014 UN271}
\shortauthors{Roth et al.}

\begin{document}


\title{The First Detection of Molecular Activity in the Largest Known Oort Cloud Comet: ALMA Imaging of C/2014 UN271 (Bernardinelli-Bernstein) at 16.6 au from the Sun}

\correspondingauthor{Nathan X. Roth}
\email{nathaniel.x.roth@nasa.gov}

\author[0000-0002-6006-9574]{Nathan X. Roth}
\affiliation{Solar System Exploration Division, Astrochemistry Laboratory Code 691, NASA Goddard Space Flight Center, 8800 Greenbelt Rd, Greenbelt, MD 20771, USA}
\affiliation{Department of Physics, American University, 4400 Massachusetts Ave NW, Washington, DC 20016, USA}
\email{nathaniel.x.roth@nasa.gov}

\author[0000-0001-7694-4129]{Stefanie N. Milam}
\affiliation{Solar System Exploration Division, Astrochemistry Laboratory Code 691, NASA Goddard Space Flight Center, 8800 Greenbelt Rd, Greenbelt, MD 20771, USA}
\email{stefanie.n.milam@nasa.gov}

\author[0000-0001-8233-2436]{Martin A. Cordiner}
\affiliation{Solar System Exploration Division, Astrochemistry Laboratory Code 691, NASA Goddard Space Flight Center, 8800 Greenbelt Rd, Greenbelt, MD 20771, USA}
\affiliation{Department of Physics, The Catholic University of America, 620 Michigan Ave., N.E. Washington, DC 20064, USA}
\email{martin.cordiner@nasa.gov}

\author{Dominique Bockelée-Morvan}
\affiliation{LIRA, Observatoire de Paris, Université PSL, CNRS, Sorbonne Université, Université Paris Cité, 5 place Jules Janssen, 92195 Meudon, France}
\email{dominique.bockelee@obspm.fr}

\author[0000-0003-2414-5370]{Nicolas Biver}
\affiliation{LIRA, Observatoire de Paris, Université PSL, CNRS, Sorbonne Université, Université Paris Cité, 5 place Jules Janssen, 92195 Meudon, France}
\email{nicolas.biver@obspm.fr}

\author[0000-0002-6702-7676]{Michael S. P. Kelley}
\affiliation{Department of Astronomy, University of Maryland, College Park, MD 20742-0001, USA}
\email{msk@astro.umd.edu}

\author[0000-0001-9479-9287]{Anthony J. Remijan}
\affiliation{National Radio Astronomy Observatory, 520 Edgemont Rd, Charlottesville, VA 22903, USA}
\email{aremijan@nrao.edu}

\author[0000-0001-6752-5109]{Steven B. Charnley}
\affiliation{Solar System Exploration Division, Astrochemistry Laboratory Code 691, NASA Goddard Space Flight Center, 8800 Greenbelt Rd, Greenbelt, MD 20771, USA}
\email{steven.b.charnley@nasa.gov}

\author[0000-0002-4043-6445]{Carrie E. Holt}
\affiliation{Las Cumbres Observatory, 6740 Cortona Drive, Suite 102, Goleta, CA 93117, USA}
\email{cholt@lco.global}

\author[0000-0001-6192-3181]{Kiernan D. Foster}
\affiliation{Department of Chemistry, University of Virginia, Charlottesville, VA, 22903, USA}
\email{cbg4vy@virginia.edu}

\author[0000-0002-1278-5998]{Joseph Chatelain}
\affiliation{Las Cumbres Observatory, 6740 Cortona Drive, Suite 102, Goleta, CA 93117, USA}
\email{jchatelain@lco.global}

\author[0000-0001-5749-1507]{Edward Gomez}
\affiliation{Las Cumbres Observatory, 6740 Cortona Drive, Suite 102, Goleta, CA 93117, USA}
\email{egomez@lco.global}

\author[0000-0002-4439-1539]{Sarah Greenstreet}
\affiliation{Department of Astronomy and the DIRAC Institute, University of Washington, 3910 15th Avenue NE, Seattle, WA 98195, USA}
\email{sarahjg@uw.edu}

\author[0000-0002-3818-7769]{Tim Lister}
\affiliation{Las Cumbres Observatory, 6740 Cortona Drive, Suite 102, Goleta, CA 93117, USA}
\email{tlister@lco.global}

\author[0000-0002-8658-5534]{Helen Usher}
\affiliation{The Open University, Walton Hall, Milton Keynes, MK7 6AA, UK}
\email{helen.usher@open.ac.uk}



\input{Abstract}

\keywords{Molecular spectroscopy (2095) ---
High resolution spectroscopy (2096) --- Radio astronomy (1338) --- Comae (271) --- Radio interferometry(1346) --- Comets (280)}


\input{Body}

\begin{acknowledgments}
This work was supported by the NASA Solar System Observations program (80NSSC24K1324: N.X.R., S.N.M., M.A.C.; 80NSSC20K0673: M.S.P.K.), the Planetary Science Division Internal Scientist Funding Program through the Fundamental Laboratory Research (FLaRe) work package (N.X.R., S.N.M., M.A.C., S.B.C.), as well as the NASA Astrobiology Institute through the Goddard Center for Astrobiology (proposal 13-13NAI7-0032; S.N.M., M.A.C., S.B.C.). It makes use of the following ALMA data: ADS/JAO.ALMA \#2023.1.00703.S and \#2011.0.00001.CAL. ALMA is a partnership of ESO (representing its member states), NSF (USA) and NINS (Japan), together with NRC (Canada), NSTC and ASIAA (Taiwan), and KASI (Republic of Korea), in cooperation with the Republic of Chile. The Joint ALMA Observatory is operated by ESO, AUI/NRAO and NAOJ. The National Radio Astronomy Observatory is a facility of the National Science Foundation operated under cooperative agreement by Associated Universities, Inc. This paper is based on observations made with the MuSCAT instruments, developed by the Astrobiology Center (ABC) in Japan, the University of Tokyo, and Las Cumbres Observatory (LCOGT). MuSCAT3 was developed with financial support by JSPS KAKENHI (JP18H05439) and JST PRESTO (JPMJPR1775), and is located at the Faulkes Telescope North on Maui, HI (USA), operated by LCOGT. MuSCAT4 was developed with financial support provided by the Heising-Simons Foundation (grant 2022-3611), JST grant number JPMJCR1761, and the ABC in Japan, and is located at the Faulkes Telescope South at Siding Spring Observatory (Australia), operated by LCOGT. Some observations were obtained by the Comet Chasers schools outreach program (www.cometchasers.org) which is funded by the UK Science and Technology Facilities Council (via the DeepSpace2DeepImpact Project),the Open University and Cardiff University.  It accesses the LCOGT telescopes through the Faulkes Telescope Project (FTPEPO2014A-004), which is partly funded by the Dill Faulkes Educational Trust, and the LCO Global Sky Partners Programme. We thank an anonymous referee for their feedback, which we feel improved the manuscript.
\end{acknowledgments}

\software{Astropy \citep{astropy:2013, astropy:2018, astropy:2022},
Astroquery \citep{Ginsburg2019},
CASA \citep{McMullin2007},
lmfit \citep{Newville2016},
Small-Bodies-Node/ice-sublimation \citep{VanSelous2021}
vis-sample \citep{Loomis2018},
Uncertainties: a Python package for calculations with uncertainties, Eric O. LEBIGOT} 

\appendix
\input{AppendixFlux}
\input{AppendixObserving}
\input{AppendixGas}

\input{AppendixActive}
\input{AppendixContinuum}


\bibliography{UN271}{}
\bibliographystyle{aasjournalv7}



\end{document}

%% file: Abstract.tex
\begin{abstract}

We report observations of comet C/2014 UN271 (Bernardinelli-Bernstein) carried out on UT 2024 March 8 and 17 at a heliocentric distance (\rh{}) of 16.6 au using the Atacama Large Millimeter/Submillimeter Array (ALMA). The CO (\Ju{}=2--1) line at 230 GHz was detected along with continuum emission from its dust coma and large ($\sim$140 km) nucleus, revealing the nature of the activity drivers and outgassing kinematics of the largest Oort cloud comet discovered to date. This work presents spectrally integrated flux maps, autocorrelation spectra, production rates, and parent scale lengths for CO and a stringent upper limit for the H$_2$CO production rate. CO outgassing displayed multiple active jets which evolved from one epoch to the next. The continuum emission was compact and spatially unresolved, and is consistent with thermal emission from the large nucleus and a tentative detection of a dust coma. Complementary optical observations provided activity context for the ALMA epochs, indicating that UN271 underwent an outburst in late February before returning to a quiescent brightness in mid--late March. These results represent the first secure detection of molecular activity reported in the literature for C/2014 UN271 and highlight the dynamic nature of this distantly active small world. 

\end{abstract}

%% file: Body.tex
\section{Introduction} \label{sec:intro}
Comets may serve as ``fossils'' of solar system formation, with the volatile composition of their nuclei reflecting the chemistry and prevailing conditions present where and when they formed \citep{Bockelee2004,Mumma2011a}. The vast majority are measured in the inner solar system during sufficient coma activity for remote sensing. Discovered through data mining of the Dark Energy Survey with activity out to \rh{} = 29 au, comet C/2014 UN271 (Bernardinelli-Bernstein; hereafter UN271) is the most distant active comet discovered to date \citep{Bernardinelli2021}. With a likely previous perihelion distance of 237 au and an orbital period millions of years long, its current perihelion passage represents its first voyage into the inner solar system \citep{Dybczynski2022}. It will reach perihelion at $q=10.9$ au, just beyond the orbit of Saturn, on 2031 January 29. ALMA and HST confirmed a large nucleus size $\sim$140 km \citep{Hui2022,Lellouch2022}. With its large size and relatively ``pristine'' preservation in the Oort cloud, UN271 may serve as a window into the composition of small Kuiper Belt Objects (KBO), which are too far from the Sun to undergo outgassing and whose compositions are being newly revealed with JWST \citep[\eg{}][]{Brown2023,Emery2024}.

Preliminary results from JWST observations near \rh{} = 18 au provided evidence of CO and CO$_2$ outgassing in UN271, but the drivers of cometary activity and the nature of CO release (\ie{} direct nucleus sublimation vs.\ coma chemistry) have not been addressed \citep{Bolin2023}. Evidence for CO$_2$ at such large \rh{} is surprising and raises the possibility of CO production through CO$_2$ photolysis. Understanding the nature and origins of activity in this distant active object requires a detailed investigation.

We report ALMA observations taken near \rh{} = 16.6 au and geocentric distance $\Delta$ = 16.9 au. We leveraged the highest spectral resolution and the most compact ALMA array configuration, aiming to discern the outgassing mechanisms of this large, distant comet. With spectral resolution orders of magnitude higher than infrared facilities, velocity-resolved spectral line profiles can be measured using ALMA, revealing coma kinematics. Furthermore, the large number of ALMA baselines sample emission over a range of angular scales. Coupled with radiative transfer modeling, such measurements can provide tight constraints on  molecular production, distinguishing direct nucleus release from production via coma chemistry \citep{Boissier2007,Cordiner2014,Roth2021a}. A key diagnostic is the parent photodissociation rate ($\beta_p$) for daughter species, which can be compared against the photodissociation rates of molecules with compatible photolysis pathways, thereby providing identifying evidence for the progenitor material.

UN271 is known to produce outbursts with relative strengths up to and around -1 mag \citep{Kelley2022,Kokotanekova2024}. Contemporaneous optical photometry with the Las Cumbres Observatory's global network of telescopes provided important context for the comet's activity state surrounding the ALMA observations. Observations in 2024 February and March provided r- and g-band images sampling UN271's dust coma and enabling lightcurve analysis alongside the ALMA spectra.

Molecular emission was identified from CO as well as continuum emission from the large nucleus and/or dust grains in the coma. Our measure of the CO (\Ju{}=2--1) transition represents the first spectroscopic detection of outgassing in UN271 reported in the literature, and the most distant detection of CO in any comet at millimeter wavelengths. A 3$\sigma$ upper limit was determined for H$_2$CO production. We report production rates and spatial maps of the detected emission. Section~\ref{sec:obs} discusses the observations and data reduction. Section~\ref{sec:results} presents our results from these data. Section~\ref{sec:modeling} discusses the kinematics and origins of CO outgassing in UN271. Section~\ref{sec:cont} analyzes the thermal continuum emission.

\section{Observations and Data Reduction} \label{sec:obs}
We conducted pre-perihelion observations toward UN271 on UT 2024 March 8 and 17 using the ALMA 12 m array with the Band 6 receiver, covering frequencies between 216 and 234 GHz in four non-contiguous spectral windows ranging from 58 MHz to 2 GHz wide. We tracked the comet position using JPL Horizons ephemerides (JPL \#60). Mean precipitable water vapor at zenith was 1.23--2.20 mm. Quasar observations were used for bandpass and phase calibration, as well as calibrating UN271's flux scale, which is crucial for an accurate comparison with previous ALMA observations \citep[\ie{}][]{Lellouch2022}. We examined the quasar fluxes and applied post-pipeline corrections to the flux calibration as detailed in Appendix~\ref{sec:fluxcal}. 

\begin{figure*}
\gridline{\fig{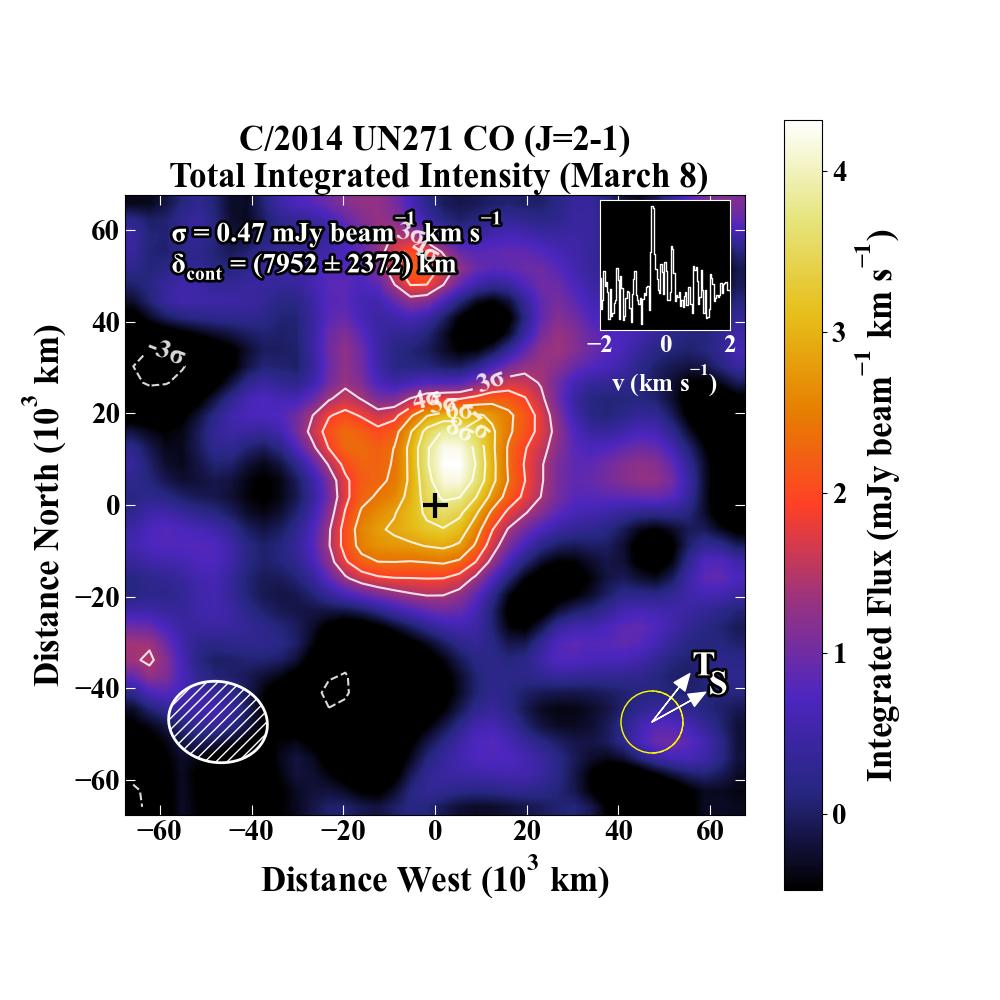}{0.45\textwidth}{(A)}
          \fig{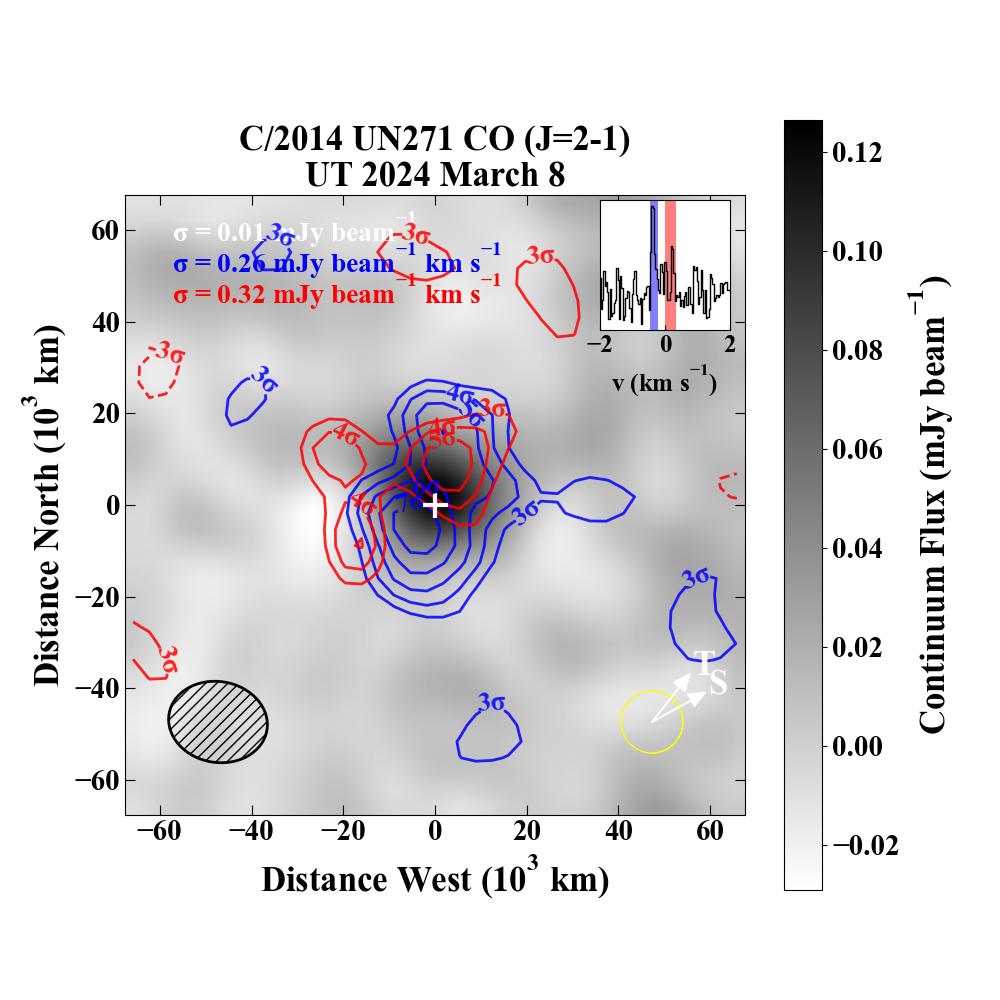}{0.45\textwidth}{(B)}
}
\gridline{\fig{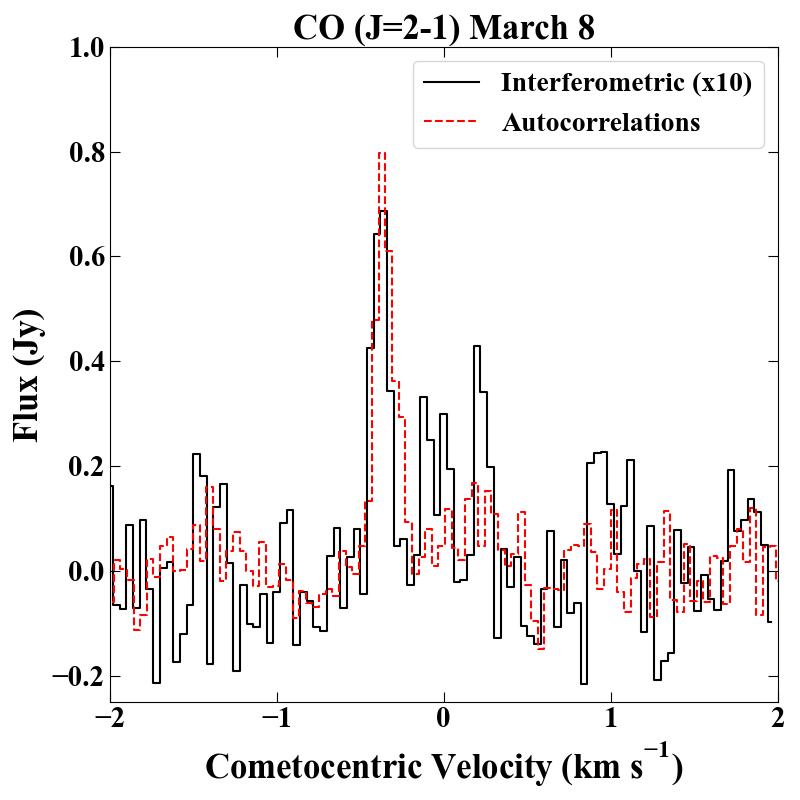}{0.45\textwidth}{(C)}
          \fig{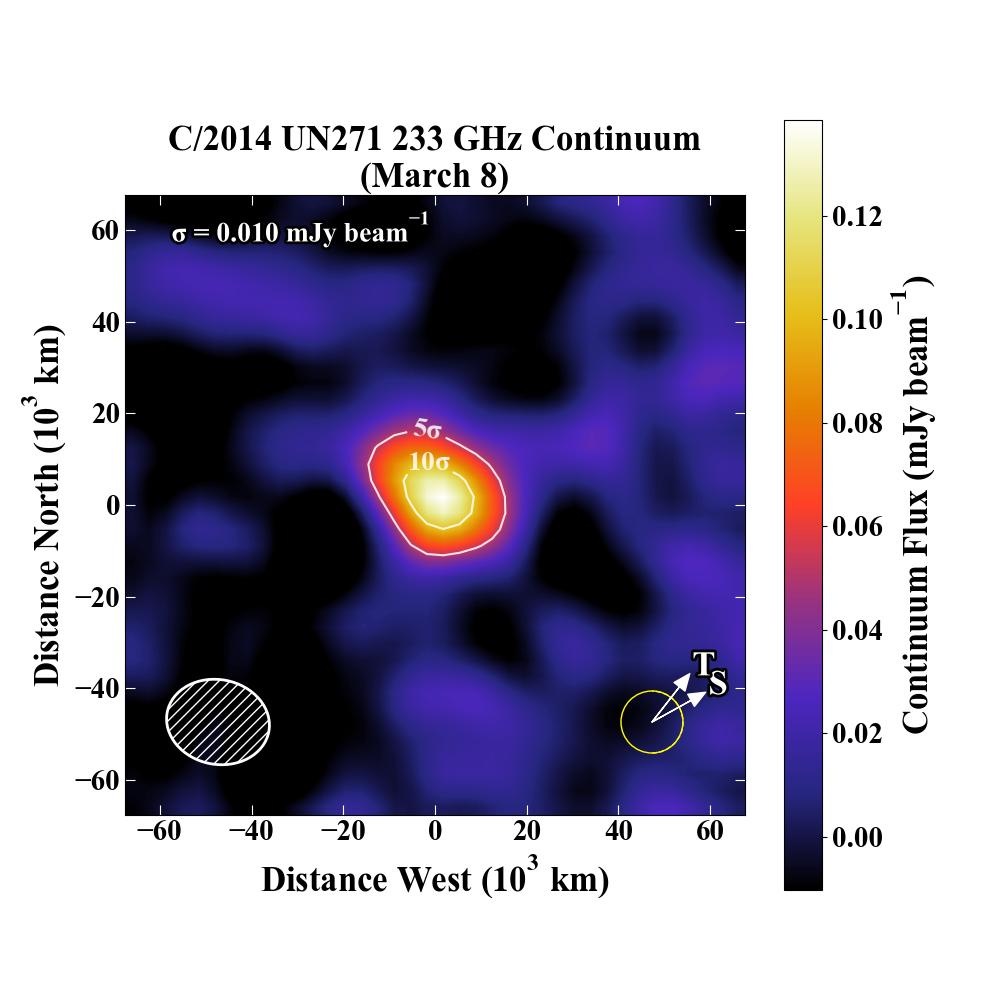}{0.45\textwidth}{(D)}
}
\caption{\textbf{(A).} Spectrally integrated flux map for CO on 2024 March 8. Contour intervals are in 1$\sigma$ increments of the rms noise, with the lowest contour being 3$\sigma$. Sizes and orientations of the synthesized beam (Table~\ref{tab:obslog}) are indicated in the lower left corner. The comet's illumination phase ($\phi \sim$ 3$\degr$), as well as the direction of the Sun and dust trail, are indicated in the lower right. The black cross indicates the peak continuum position. $\sigma$ gives the RMS noise and $\delta$\subs{cont} the separation between the location of peak continuum and peak gas emission. A spectral line profile taken at the continuum peak in a 10$\arcsec$ diameter aperture is shown. \textbf{(B).} Continuum flux map with contours for the spectrally integrated red and blue CO components overlaid separately. The extracted spectral line profile shows the integration region for the blue and red maps across the line. \textbf{(C).} Comparison of autocorrelation and interferometric CO spectra. \textbf{(D).} Continuum flux map in UN271. Contours are in 5$\sigma$ increments of the rms noise, with the lowest contour being 5$\sigma$.
\label{fig:maps}}
\end{figure*}

The spatial scale (range in semi-minor and semi-major axes of the synthesized beam) was 1$\farcs$48 -- 2$\farcs$55 and the frequency resolution was 244 kHz for the H$_2$CO spectral line window and 31 kHz for the CO spectral window, resulting in velocity resolution of 0.04--0.33 km s$^{-1}$. The data were flagged, calibrated, and imaged using standard routines in Common Astronomy Software Applications (CASA) package version 6.6.1 \citep{McMullin2007}. We deconvolved the point-spread function with the Högbom algorithm, using natural visibility weighting, a 10$\arcsec$ diameter mask centered on the peak continuum position, and a flux threshold of twice the rms noise. The deconvolved images were then convolved with the synthesized beam and corrected for the (Gaussian) response of the ALMA primary beam. We transformed the images from astrometric coordinates to projected cometocentric distances, with the location of the peak continuum flux chosen as the origin, which was in good agreement with the predicted ephemeris position. 

Optical observations of the comet were taken between 2024 February 03 and March 24 with the Las Cumbres Observatory network of telescopes, primarily from 1-m telescopes at Cerro Tololo International Observatory, Chile, but also from 1-m telescopes at the South African Optical Observatory, Sutherland, South Africa and the 2-m Faulkes Telescope South at Siding Spring Observatory, Coonabarabran, Australia.  Images of the comet were taken through SDSS $r'$ and $g'$ filters.  Observation sequences with the 1-m telescopes used the Sinistro cameras (4k$\times$4k CCDs, 0\farcs39 per pixel) and consisted of 2$\times$180~s images per filter. The Faulkes Telescope South observations used the MuSCAT 4 camera, which provides simultaneous observations through four filters \citep[2k$\times$2k CCDs, 0\farcs27 per pixel;][]{narita20}, but only the $g'$ and $r'$ data (4$\times$120~s each) are analyzed here.  All telescopes tracked the comet at the ephemeris rate.  All data were processed with the BANZAI pipeline \citep{mccully18}, and photometrically calibrated to PS1 $r$- and $g$-band photometry using background stars and a color correction \citep[see]{lister22}.  We adopt a $g-r$ color of 0.46~mag, based on a broader dataset from the LCO telescopes (Kokotanekova et al., in preparation).  The brightness of the comet was measured in 6\arcsec{} radius apertures (73,000 to 74,000~km), and data affected by nearby stars and poor image quality (FWHM $>$3\arcsec), were removed from the analysis.  Observing circumstances for ALMA and LCO along with LCO photometry are provided in Appendix~\ref{sec:obslog}.

\subsection{Autocorrelation Spectra}\label{subsec:autocorr}
We extracted autocorrelation spectra at the expected frequency for CO (\Ju{} = 2--1) as well as for the H$_2$CO ($J_{Ka,Kc}$ = $3_{0,3}$--$2_{0,2}$) transition (detected in neither cross- nor auto-correlations) to provide total power spectra of extended flux that may have been resolved out by the 12 m array. These autocorrelation spectra are collected for each antenna simultaneously with the interferometry, but are flagged by the ALMA pipeline. We recovered, extracted, and calibrated the autocorrelation spectra as in \cite{Cordiner2023}. The spectra were converted to a flux scale (Jy) using the beam sizes and aperture efficiencies in the ALMA Technical Handbook \citep{Cortes2024}.

\section{Results}\label{sec:results}
\subsection{ALMA Results}\label{subsec:alma_results}
We securely detected CO (\Ju{}=2--1) and continuum emission on both dates, and derived sensitive upper limits for H$_2$CO production. CO was detected in interferometric and autocorrelation spectra. Figure~\ref{fig:maps} shows spectrally integrated flux maps for CO and continuum in UN271 on March 8, as well as a comparison of the CO interferometric and autocorrelation spectra. Spectra and maps for March 17 are in Figures~\ref{fig:maps17} and~\ref{fig:cont_compare}. Compared to March 8, CO emission on March 17 was more broadly extended in the sunward direction but detected at lower statistical significance. We applied a 2$\arcsec$ Gaussian taper in the $uv$-plane to the March 17 CO image to improve the signal-to-noise ratio (S/N). This Gaussian taper down-weights signal from longer baselines, increasing surface brightness sensitivity at the cost of decreased angular resolution. Differences are evident between molecular and continuum emission. The continuum was relatively compact about the nucleus at the 5$\sigma$ level, whereas molecular emission extended to greater distances. The CO spatial distribution evolved between epochs although the lower image S/N on March 17 complicates interpretation. Whereas the CO spectral line profiles in the autocorrelations remained consistent, there was significant evolution in the interferometric spectra.

Offsets are evident between the continuum peak (taken to be the nucleus) and molecular emission peak and are quantified in the maps. The offsets are smaller than the FWHM (full width at half maximum) of the synthesized beam (Table~\ref{tab:obslog}) and measured with 2--3$\sigma$ confidence. Since continuum and molecular emission were simultaneously measured, the uncertainty in these offsets is given by the FWHM of the synthesized beam divided by the S/N of the weaker of the two. The offset nominally increases between March 8 and 17, although the two are in formal agreement.

Further insights into CO arise from the integrated flux maps of the blue and red components. Figure~\ref{fig:maps}B shows that on March 8, there was a clear contribution from both the blue and red features along the projected sunward (northwest) direction. There was also a strong blue feature and a weaker red feature along an opposite position angle, but the small solar phase angle complicates interpretation. Although the S/N is lower on March 17, the sunward blue feature remained and dominated the emission (Figure~\ref{fig:maps17}C). The weak southwestern red feature was still detected, but the sunward red feature and southwestern blue feature were not.

\subsection{LCO Results}\label{subsec:lco_results}
\begin{figure}
\plotone{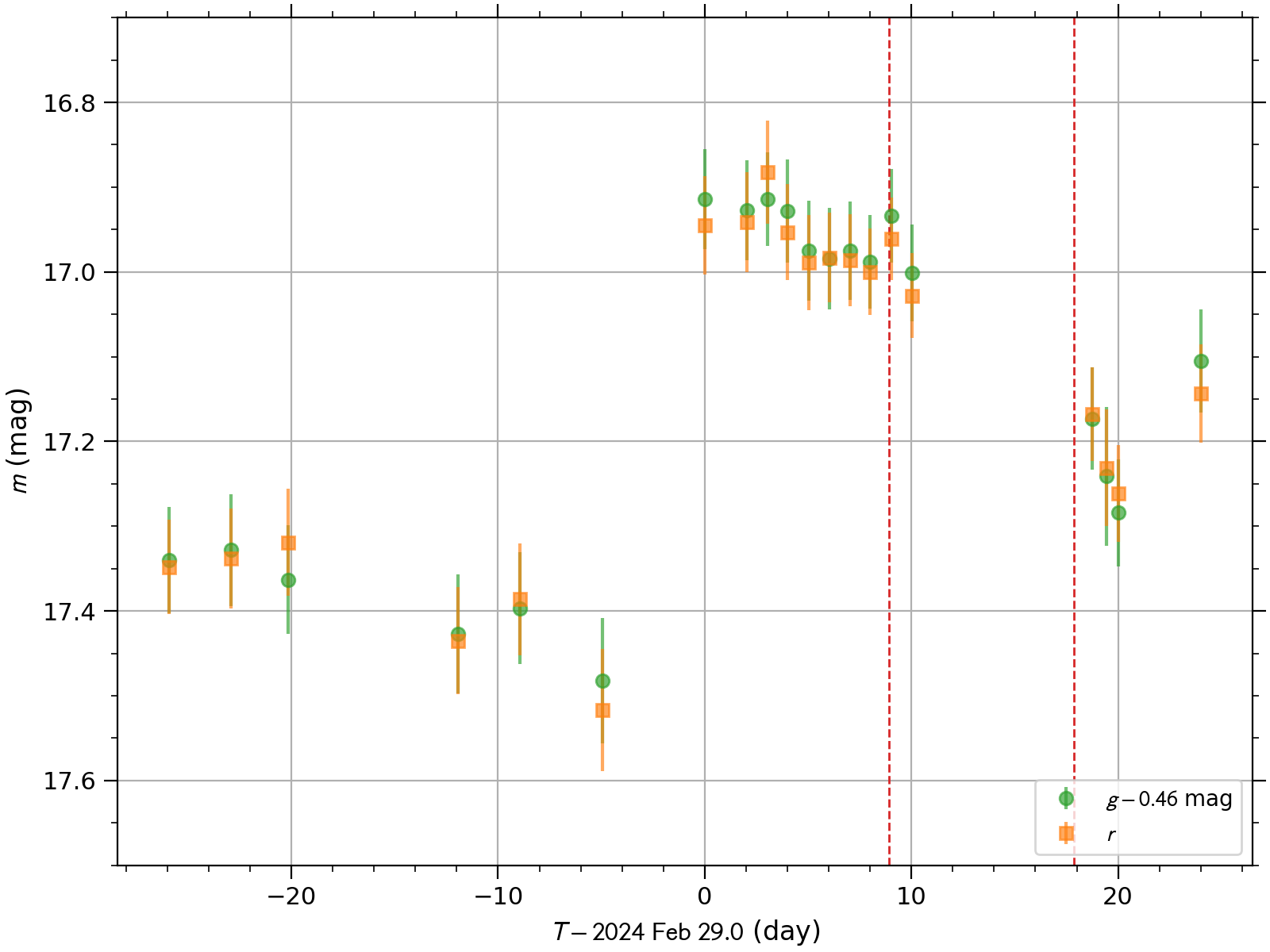}
\caption{Optical lightcurve of comet C/2014 UN271 versus time from
2024 February 29.0 (UTC) based on $g$-band photometry (circles) and
$r$-band photometry (squares).  The $g$-band data has been scaled by
the color of the comet (--0.46~mag) to match the $r$-band photometry.
Vertical dashed lines mark the times of the ALMA observations.
\label{fig:lco}}
\end{figure}

The LCO lightcurves provide an optical counterpart to examine UN271's overall activity state. At least one outburst occurred in 2024 February/March.  The comet's apparent brightness (Fig.~\ref{fig:lco}) was near-constant in the month of February, averaging $r=17.38\pm0.03$~mag.  Between February 24 and 29, the comet brightened by $-0.44\pm0.06$~mag (Table~\ref{tab:lco}).  The comet gradually nearly returned to its February brightness before possibly brightening again between March 20 and 24.  The first brightening has the appearance of a cometary outburst, where a significant amount of material (dust and ice) is ejected from from the nucleus on a short timescale, and that material slowly disperses with time.  The second brightening may also be an outburst, but its interpretation depends on whether or not the comet's brightness would have further declined after March 20.  Taking 2024 February 26.53 (UTC) as the nominal outburst time (the center point of the February 24 -- 29 observational gap), the ALMA observations occurred approximately 11.4 and 20.4 days after the event, with a full-range uncertainty of $\pm$2.5~days.

The shape of the post-outburst lightcurve is more rounded than events described previously. \cite{Kelley2022} showed that the 2021 lightcurve events could be approximated with exponential functions in magnitude units.  Similarly shaped outburst lightcurves may be seen in the literature \citep[e.g.,][]{trigo-rodriguez08,ishiguro16,kelley21,gillan24}.  Generally the shape of an outburst lightcurve can be affected by photometric aperture size, the 3D dynamical expansion of the material relative to the observer, grain fragmentation, grain sublimation, and whether or not the outburst location's activity changed.  Specific to the February outburst of UN271, the slow decay might be due to a slow expansion speed (as projected on the sky plane), grain fragmentation, and/or additional (short-lived) activity from the outburst location.

\section{Modeling and Interpretation}\label{sec:modeling}

\subsection{Analysis of CO Emission and H$_2$CO Upper Limits}\label{subsec:kinematics}
The CO line profiles were strongly asymmetric with a dominant blue component, ruling out isotropic outgassing. Although the CO autocorrelation spectra were similar on March 8 and 17, the interferometric spectra showed significant differences between epochs, as well as to the autocorrelations (Figures~\ref{fig:maps}C and~\ref{fig:maps17}D). The cometocentric velocity of the red peak shifted significantly in the interferometric spectra, and the asymmetry between the blue and red peaks was larger in the autocorrelations than cross-correlations. However, all spectra share a consistent velocity for the blue peak. A comparison of UN271's astrometric position against \textit{Planck} CO maps \citep{Ghosh2024} and SIMBAD queries \citep{Wenger2000} confirmed there were no known background sources within 100$\arcsec$ (the ALMA primary beam is $\sim25\arcsec$ at 230 GHz). Thus, the complex and varying line shapes should be attributed to CO outgassing from UN271.

Given the $\sim$10 years photolysis lifetime of CO at the \rh{} of UN271 along with the disconnect in morphology between the autocorrelation and interferometric spectra, the interferometry sampled more nearly thermalized CO proximal to the nucleus ($\sim$20,000 km) and indicative of its activity. In contrast, the autocorrelations measured an extended CO coma ($\sim$250,000 km) in fluorescence equilibrium, including material ejected during outbursts, dominated by the consistent blue component. Given the blue component expansion speed ($\sim$0.38 \kms{} from the blue peak velocity), it would take CO $\sim$9 days to travel from the 1.7$\arcsec$ synthesized beam to the 25$\arcsec$ edge of the FOV, making a comparison of the March 8 interferometry more appropriate with the March 17 autocorrelations. Although previous works have modeled the autocorrelation and interferometric spectra simultaneously \citep[\eg{}][]{Roth2021a,Cordiner2023}, we elected to treat only the interferometric spectra for these reasons.

We modeled molecular line emission using the SUBLIMED three-dimensional radiative transfer code for cometary atmospheres \citep{Cordiner2022}, including a full non-LTE treatment of coma gases, collisions with CO, and pumping by solar radiation. We used the escape probability formalism \citep{Bockelee1987} to address opacity for the ultra-cold CO emission, along with a time-dependent integration of the energy level population equations. We used explicitly calculated CO-CO collisional rates \citep{Cordiner2022}. We are unaware of published H$_2$CO-CO collisional rates, so we calculated them using the thermalization approximation \citep{Crovisier1987,Biver1999,Bockelee-Morvan2012}, using an average collisional cross-section with CO of $5\times10^{-14}$ cm$^{-2}$. Photodissociation rates were adopted from \cite{Hrodmarsson2023}.

We followed methods applied to CO-rich comets C/2016 R2 (PanSTARRS) and 29P/Schwassmann-Wachmann 1 \citep{Cordiner2022,Roth2023}, dividing the coma into two outgassing regions, $R_1$ and $R_2$, each with independent molecular production rates ($Q_1$,$Q_2$), gas expansion velocities ($v_1$,$v_2$), and parent scale lengths ($L_{p1},L_{p2}$). $L_p$ is the distance from the nucleus at which CO formed ($L_p$ = 0 km is direct nucleus release) and is related to the parent photodissociation rate as $L_p = v/\beta_p$. The narrow width of the blue component and the line velocity shift are consistent with CO outflow from the subsolar point instead of uniform outflow. This is the simplest  model capable of fitting the asymmetric line profiles. Details regarding our modeling and parameter optimization approach are in Appendix~\ref{sec:fourier}.

\begin{figure*}
\gridline{\fig{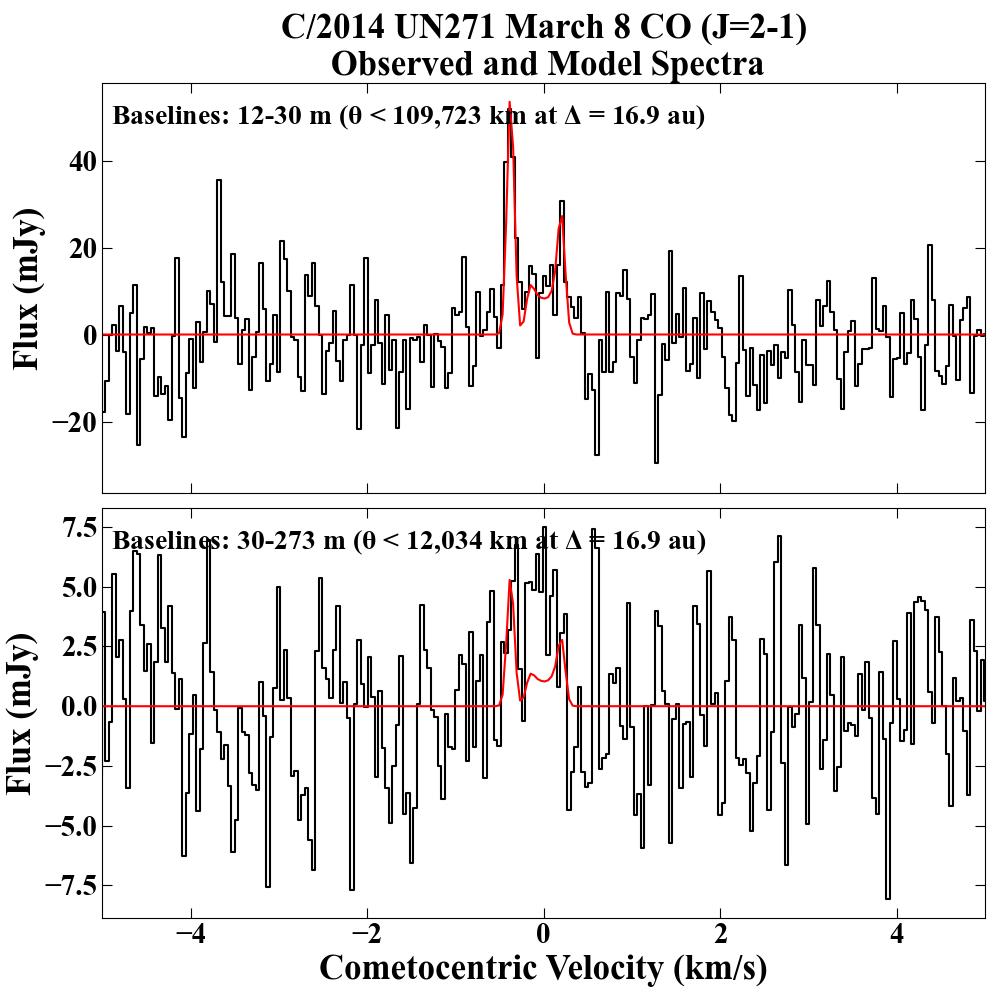}{0.45\textwidth}{(A)}
          \fig{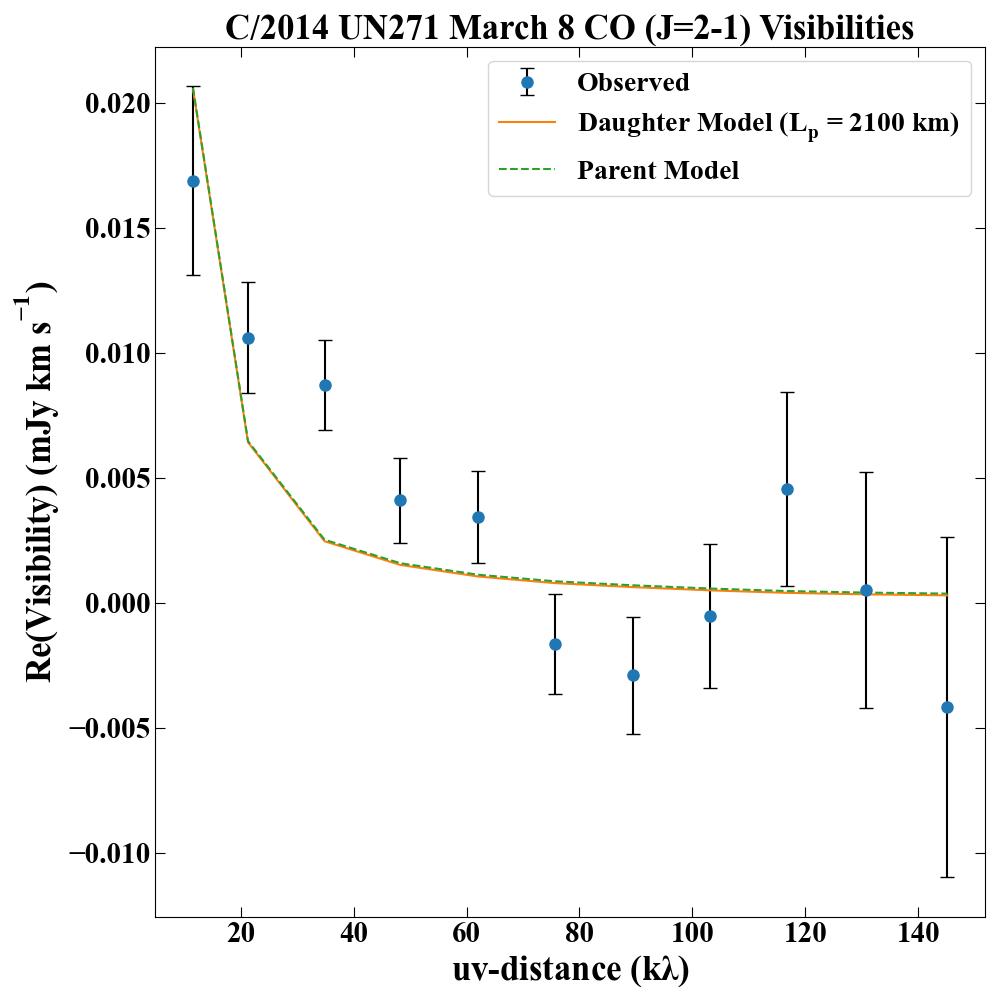}{0.45\textwidth}{(B)}
          }
\gridline{\fig{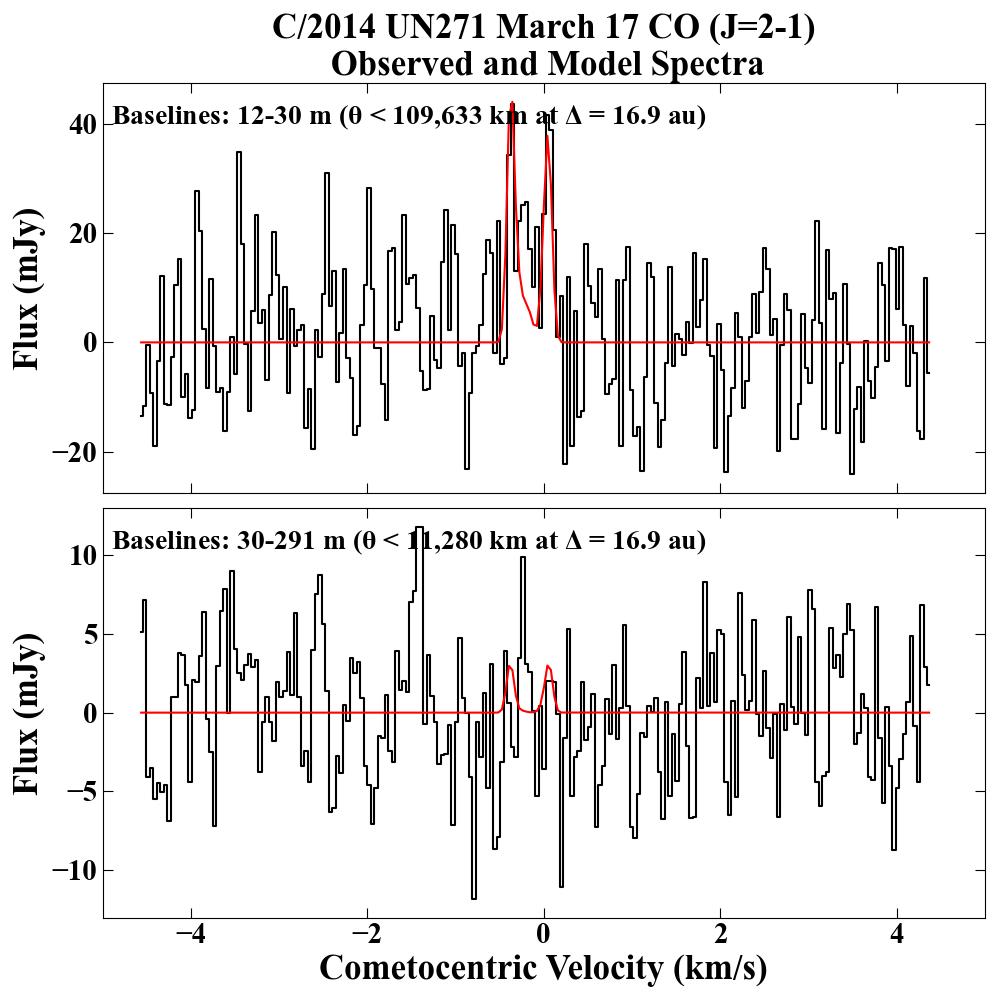}{0.45\textwidth}{(C)}
          \fig{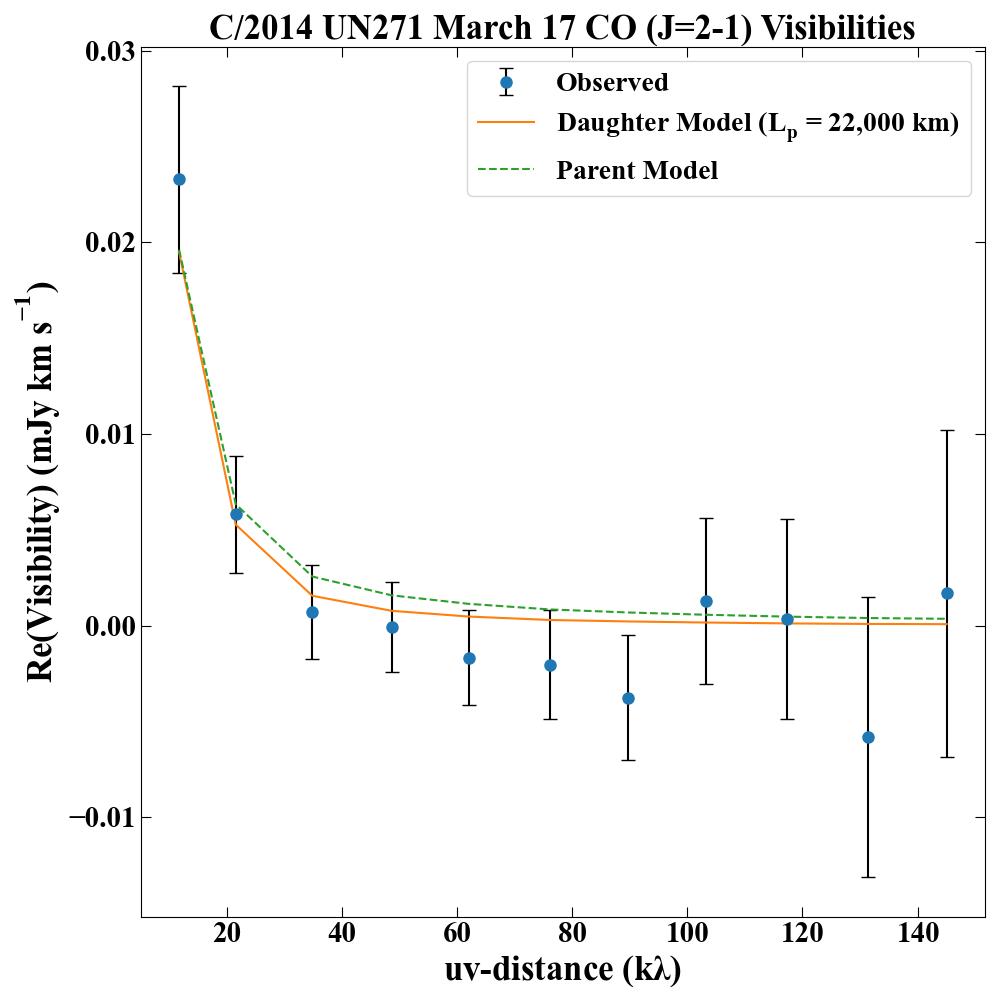}{0.45\textwidth}{(D)}
          }
\caption{\textbf{(A).} CO spectrum in UN271 on March 8 (black), with each panel representing spectra extracted from differing baseline ranges (angular scales). The spectra are displayed with a frequency resolution of 61 kHz (velocity resolution 0.08 \kms{}). The best-fit model is overplotted in red. \textbf{(B).} Real part of the observed visibility amplitude as a function of projected baseline length for CO on March 8. The upper limit daughter model, as well as a parent model, are overplotted for comparison. The $uv$ distance has been plotted in units of k$\lambda$. \textbf{(C--D).} Plots for March 17 with traces and labels as in panels (A)--(B).
\label{fig:fits}}
\end{figure*}

Our best-fit models are summarized in Table~\ref{tab:comp} and Figure~\ref{fig:fits}, with $Q$(CO) ranging from (3.5--4.7)$\times10^{27}$ s$^{-1}$ for direct nucleus release. If CO in UN271 were produced as a daughter species, our 1$\sigma$ upper limit on $L_p$ ($<$2,500 km for March 8) would correspond to $\beta_p$(\rh{}=1 au)$> 4.29\times10^{-2}$ s$^{-1}$ (where $\beta$(1 au)=$\beta$(\rh{})\rh{}$^2$). No molecule detected in comets to date with photolysis pathways to CO has a such a photodissociation rate \citep{Hrodmarsson2023}, nor sufficient abundance to reproduce our CO flux via daughter production. We interpret our results as evidence of direct nucleus CO sublimation and posit it is a key activity driver in UN271. 

The lower S/N for March 17 provides a less stringent 1$\sigma$ upper limit on $L_p$ ($<$22,000 km), corresponding to $\beta_p$(\rh{}=1 au)$> 4.80\times10^{-3}$ s$^{-1}$. Similar to March 8, no known cometary molecule could serve as a CO parent with these characteristics.

\subsubsection{Interpretation of CO Spectra in C/2014 UN271}\label{subsubsec:interpretation}
The CO spectra and maps measured in UN271 with ALMA are complex, obtained at \rh{} seldom probed by remote sensing. Perhaps the most apt comparison is C/1995 O1 (Hale-Bopp). Our $Q$(CO) at 16.6 au is similar to that measured in Hale-Bopp at \rh{} = 14 au, and if scaled with insolation-driven dependence, $Q$(CO) in UN271 at \rh{} = 1 au would be $\sim9\times10^{29}$ s$^{-1}$, consistent with Hale-Bopp near perihelion \citep{Biver2002}. However, Hale-Bopp's nucleus size was nearly half UN271's \citep{Weaver1997} and its CO/H$_2$O abundance exceptionally high, indicating that the near-surface layers of UN271 are less CO-rich or the nucleus active fraction is smaller. Owing to its large perihelion distance ($q=10.9$ au) far outside the \rh{} = 2-3 au region where H$_2$O begins to vigorously sublime, the CO/H$_2$O abundance in UN271 cannot be directly measured during the current perihelion passage for comparison with Hale-Bopp. 

However, \cite{Harrington2022} examined the relationship between $Q$(CO) and nucleus diameter ($D$) as a function of \rh{} for comets and Centaurs. Adding $Q$(CO)/$D^2$ for UN271 at \rh{} = 16.6 au to their Figure 11 is consistent with values measured in Centaurs, and the extrapolated value for $Q$(CO)/$D^2$ at \rh{} = 1 au falls in the middle of the range. Similarly, the CO active fractions can be calculated following the model of \cite{Cowan1979} and the Small-Bodies-Node/ice-sublimation code \citep{VanSelous2021}, yielding 0.07\% for UN271 at \rh{} = 16.6 au and 0.22\% for Hale-Bopp at \rh{} = 14 au (Appendix~\ref{sec:active}).

CO on March 8 (Figure~\ref{fig:maps}B) showed a dual-jet structure, with blueshifted CO oriented along diametrically opposed position angles ($\psi\sim120\degr,300\degr$). Our two-component radiative transfer model is a simplification of this geometry; however, due to optical depth and coma dynamical effects, incorporating additional jet features into the radiative transfer modeling scheme is an undertaking beyond the scope of this work.

Although at lower signal-to-noise, outgassing on March 17 was different. The peak CO intensity was more offset from the nucleus than on March 8. The blue jet at $\psi\sim300\degr$ was not detected, whether due to cessation of activity or instead rotation out of view is unclear. Although we take the 0.066 \kms{} red peak at face value for determining $v_2$, it is possibly a projection effect if the missing blue jet had rotated into the sky plane in the anti-sunward hemisphere. Alternatively, there may be significant CO production occurring from the remnants of the outburst documented by LCO, which indicated overall activity significantly decreased between March 8 and 17.  Quantitative modeling of these alternative explanations is the subject of a future work.

For all of UN271's variability, the consistency of the expansion speed for the blueshifted CO suggests it is an inherent feature of the nucleus. The gas expansion speed can be related to the surface temperature at the subsolar point. \cite{Crovisier1995} found that  the surface temperature was 100 K for $v = 0.48$ \kms{} in CO-dominated 29P/Schwassmann-Wachmann 1 at \rh{}$\sim6$ au. Following \cite{Crifo1997}, the terminal gas speed can be calculated as $v=\sqrt{(\gamma+1)(\gamma k_BT)/(m(\gamma-1))}$ where $\gamma = 1.4$ is the ratio of specific heats for CO and $m$ the CO molecular mass. Our CO expansion speeds (Table~\ref{tab:comp}) give $T$ = $(60\pm2)$ K and $(60\pm3)$ K on March 8 and 17, respectively. These are lower than the subsolar surface temperature calculated by NEATM modeling ($95\pm9$ K; Appendix~\ref{sec:cont_analysis}). 

\subsubsection{H$_2$CO Upper Limits}
We extracted H$_2$CO spectra at the nucleus position. We calculated $3\sigma$ upper limits on the integrated intensity $\int T_B \mathrm{d}v <12.5$ mK km s$^{-1}$ and $<9.9$ mK km s$^{-1}$ on March 8 and 17, respectively. We assumed direct nucleus release of H$_2$CO. Although H$_2$CO is well-established as a distributed source species produced in the coma \citep[e.g.,][]{Biver1999,Cordiner2014} there are no constraints on the expected scale length of its progenitor at \rh{}$\sim16$ au. These correspond to 3$\sigma$ upper limits $Q$(H$_2$CO)/$Q$(CO) $<$ 3.4\% and $<$3.8\% on March 8 and 17, indicating that H$_2$CO was not a significant coma component \citep[where H$_2$CO/CO ranges from 1--81\% among comets measured; see Fig. 7B of][]{Roth2023}.

\begin{deluxetable*}{ccccccc}
\tablenum{1}
\tablecaption{Molecular Emission Analysis in C/2014 UN271 \label{tab:comp}}
\tablewidth{0pt}
\tablehead{
\colhead{Transition} & \colhead{$v_1$\sups{a}} & \colhead{$v_2$\sups{b}} & \colhead{$\gamma$\sups{c}} & \colhead{$Q_1$/$Q_2$\sups{d}} & \colhead{$Q$\sups{e}} &  \colhead{\lp{}\sups{f}} \\
\colhead{} & \colhead{(\kms{})} & \colhead{(\kms{})} & \colhead{($^{\circ}$)} & \colhead{} & \colhead{($10^{27}$ s$^{-1}$)} & \colhead{(km)}
}
\startdata
\multicolumn{7}{c}{Molecular Production and Kinematics} \\
\hline
\multicolumn{7}{c}{2024 March 8, \rh{}=16.61 au, $\Delta$=16.91 au, $\phi$\subs{STO}=$3.2\degr$, $\psi_{\sun} = 119.5\degr$} \\
CO & 0.386 $\pm$ 0.006 & 0.212 $\pm$ 0.014 & 27 $\pm$ 5 & 6.3 $\pm$ 1.4 & 3.44 $\pm$ 0.55 & $<$ 2500 \\
    & 0.385 $\pm$ 0.006 & 0.213 $\pm$ 0.014 & (27) & 6.2 $\pm$ 1.4 & 3.47 $\pm$ 0.55 & (0) \\
H$_2$CO  & (0.386) & (0.212) & (27) & (6.2) & $<0.12$ $(3\sigma)$ & (0) \\ 
\hline
\multicolumn{7}{c}{2024 March 17, \rh{}=16.58 au, $\Delta$=16.87 au, $\phi$\subs{STO}=$3.2\degr$, $\psi_{\sun} = 128.1\degr$} \\
CO & 0.386 $\pm$ 0.011 & 0.066 $\pm$ 0.016 & 73 $\pm$ 7 & 31 $\pm$ 14 & 5.33 $\pm$ 0.88 & $<$22,000 \\
   & 0.388 $\pm$ 0.012 & 0.066 $\pm$ 0.016 & (73) & 28 $\pm$ 13 & 4.73 $\pm$ 0.78 & (0) \\
\enddata  
\tablecomments{\sups{a} Gas expansion speed in the sunward jet ($R_1$). \sups{b} Gas expansion speed in the ambient coma ($R_2$). \sups{c} Half-opening angle of the sunward jet. \sups{d} Ratio of production rates in the jet vs.\ the ambient coma ($R_1$ / $R_2$). \sups{e} Global molecular production rate. \sups{f} Parent scale length. 
}
\end{deluxetable*}

\section{Analysis of Continuum Emission}\label{sec:cont}
\cite{Lellouch2022} characterized thermal emission from UN271's large nucleus at 233 GHz using ALMA, providing a nucleus diameter 137 $\pm$ 17 km and a geometric albedo $p_V$ = (5.3 $\pm$ 1.2)\%. We assumed these values, including their adopted beaming factor $\eta$ = 1.175 $\pm$ 0.42, radio emissivity $\epsilon_r$ = 0.70 $\pm$ 0.13, and bolometric emissivity $\epsilon_b$ = 0.90 $\pm$ 0.06. 

\begin{figure*}
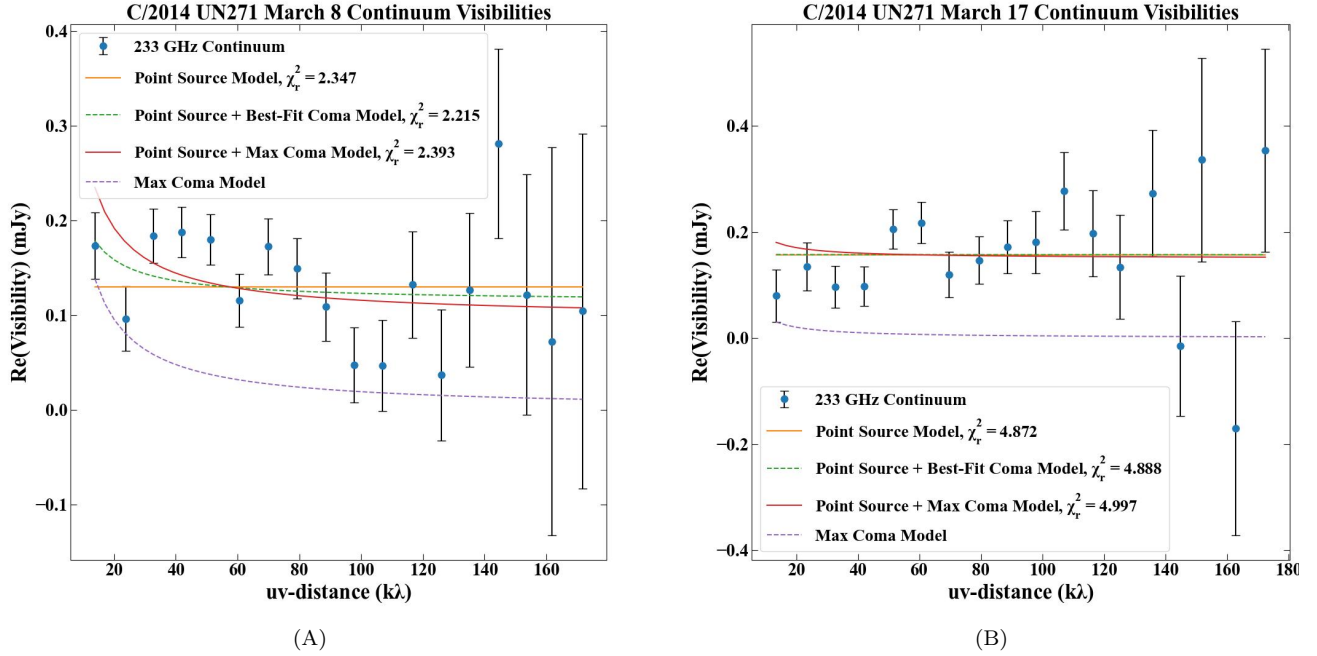

\gridline{\fig{UN271-Mar8-Continuum-Visibilities-Index.jpg}{0.45\textwidth}{(A)}
          \fig{UN271-Mar17-Continuum-Visibilities-Index.jpg}{0.45\textwidth}{(B)}
          }
\caption{\textbf{(A--B).} Continuum visibilities on March 8 and March 17, with various nucleus and coma models overplotted. The $uv$ distance has been plotted in units of k$\lambda$.
\label{fig:cont_fits}}
\end{figure*}

For ease of comparison, we scaled our 216 GHz and 232 GHz visibilities to 233 GHz using their derived spectral index ($\alpha = 1.93$).  Figure~\ref{fig:cont_fits} shows our continuum visibilities as a function of $uv$ distance. On both dates the visibility curves were flat within uncertainty, ruling out a spatially extended dust coma and instead compatible with (1) a compact, spatially unresolved coma along with the nucleus or (2) solely the nucleus. We followed their methods by performing visibility fitting using (1) a point-source model and (2) a combination of a point-source model and a dust coma which follows a 1/$\rho$ distribution. Details of our approach and a comparison of best-fit models are in Appendix~\ref{sec:cont_analysis}.

We calculated nucleus sizes using $T_{\mathrm{ss}}$ calculated by the NEATM as well as from the CO expansion speed, with nominal diameters ranging from 101 km -- 151 km. Overall, our values are in formal agreement with \cite{Lellouch2022} within 2$\sigma$. Based on $\chi^2$ analysis, our March 8 data are best fit by a 101--129 km nucleus and a dust coma on the order of $0.4-1.2\times10^{11}$ kg depending on the dust optical properties, whereas the March 17 data are best explained by a $118-151$ km nucleus with only $3\sigma$ upper limits from a coma. Likewise, we calculated a maximum contribution from any dust coma using the flux measured on the shortest baselines. Even then, the nucleus accounts for the majority of the emission at the $0.097-0.150$ mJy level on March 8 and 17, respectively. Our results support a large nucleus, but our coma dust mass for March 8 is significantly larger than the $10^7-10^8$ kg masses produced on short timescales during prior outbursts \citep{Kelley2022}; however, differing formalisms are used in each study. Indeed, optical and radio estimates of dust masses from the outburst of comet 17P/Holmes differed by a factor of 10$^4$ owing to model parameters, with the larger dust mass implied by radio measurements eventually confirmed through observations of the remnant dust cloud \citep{Ishiguro2016}. In any case, the presence of a compact dust coma on March 8 that was not detected on March 17 is consistent with the LCO lightcurve trends.

\section{Conclusion and Future Directions}\label{sec:conclusion}
The ALMA telescope array has enabled the characterization of the molecular spatial distributions, kinematics, and production mechanisms in several comets in the inner Solar System. This work adds distant active comet C/2014 UN271 to the literature. Our measurements revealed an evolving and dynamic small world, with complex outgassing where CO is a principal component. The kinematics and production mechanisms for CO are qualitatively consistent with UN271's repeated outbursts and may be indicative of activity originating from multiple sites on its large nucleus.

This work provides a measure of molecular activity for UN271 near \rh{} = 16 au, far in advance of its perihelion. As UN271 continues to approach the Sun, additional volatiles should be expected to activate, revealing the primitive chemistry preserved within. Having reached \rh{}$\sim$15 au on 2025 June 1, CH$_4$ outgassing may be recently activated based on equilibrium temperature considerations \citep[See Figure 1.2 of][]{Crovisier2007}. Next may come C$_2$H$_6$ near 12.5 au (2027 November), followed by H$_2$S and C$_2$H$_2$ near 11 au (2030 March). With its perihelion distance $q=10.9$ au, the least volatile species which may activate include H$_2$CO, CO$_2$, and potentially NH$_3$ (2031 January), whereas the prospects for HCN and CH$_3$OH are at best uncertain, and H$_2$O is  unlikely to vigorously sublime. However, UN271 is a highly complex object, and much remains to be seen regarding its behavior near perihelion. Collectively, this study could only be performed with the high sensitivity and angular resolution of ALMA, and will serve as the basis for future ALMA studies of comets at increasingly large \rh{}.

%% file: AppendixFlux.tex
\section{Flux Calibration} \label{sec:fluxcal}
Accurate flux calibration for this study is critical, particularly for comparison of our continuum fluxes against the values of \cite{Lellouch2022}. We examined the measured quasar fluxes in the ALMA Calibrator Source catalog for the flux calibrators in our study (J0334-4008 and J0538-4405 on March 8, J0519-4546 on March 17) within an 80 day window to either side of our observations. All were monitored with a high cadence at Band 3 (91.5 GHz and 103.5 GHz), with less frequent monitoring at Band 6 (233 GHz) and Band 7 (337 GHz and 343 GHz). Although all were measured at Band 6 within 20 days of our study, none had measurements on our observation date.

\begin{figure*}
\gridline{\fig{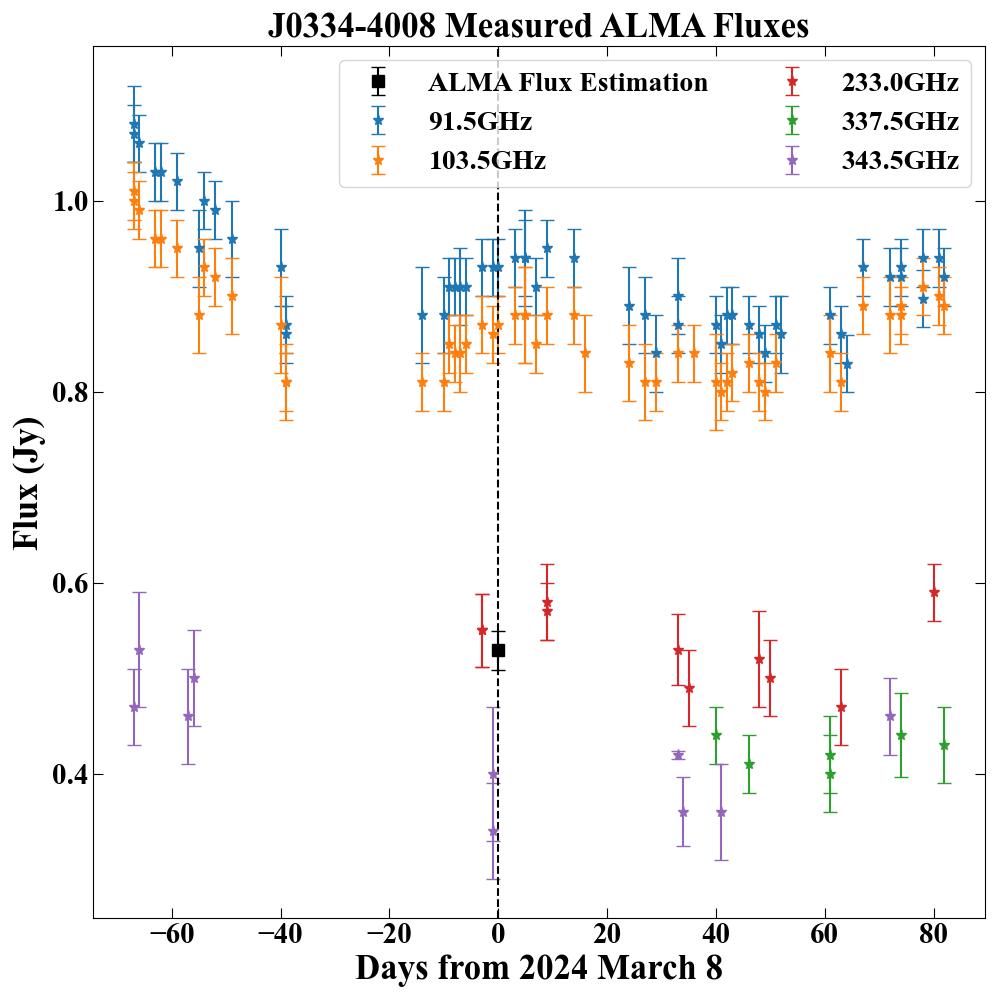}{0.45\textwidth}{(A)}
          \fig{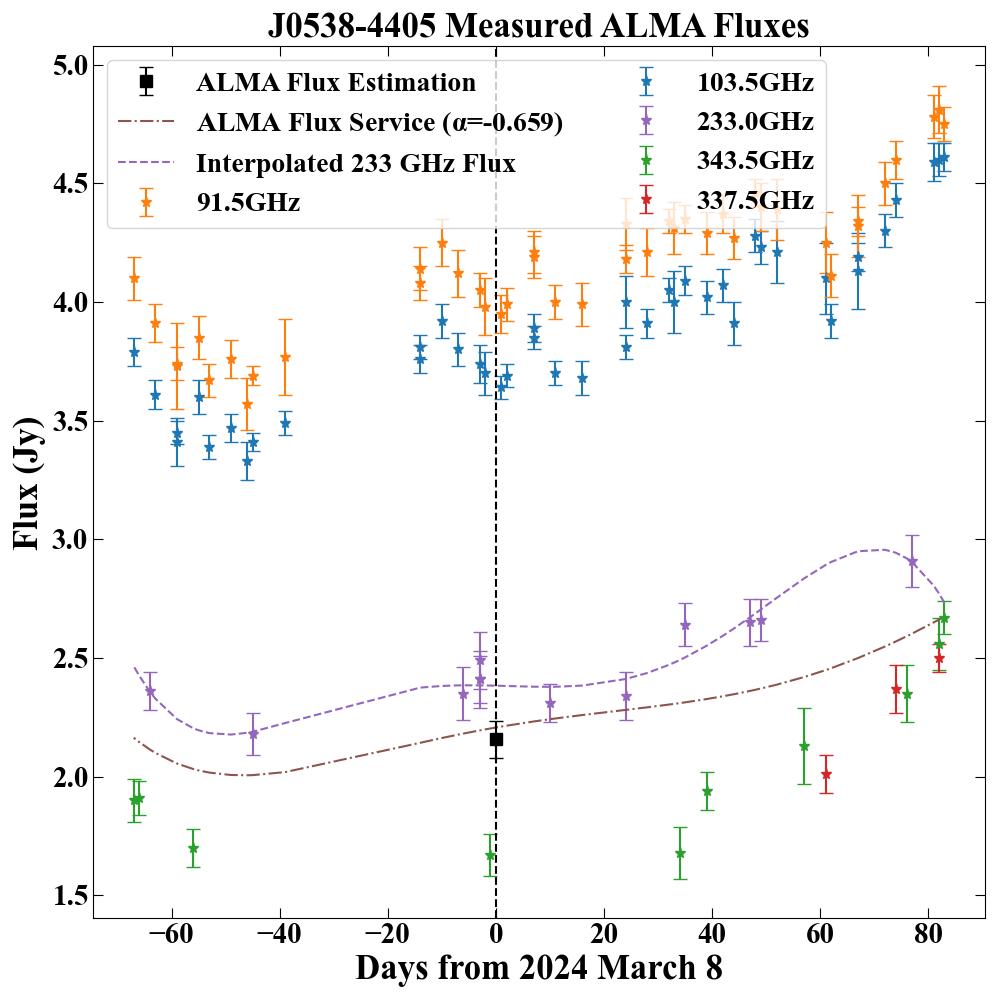}{0.45\textwidth}{(B)}
	}
\gridline{\fig{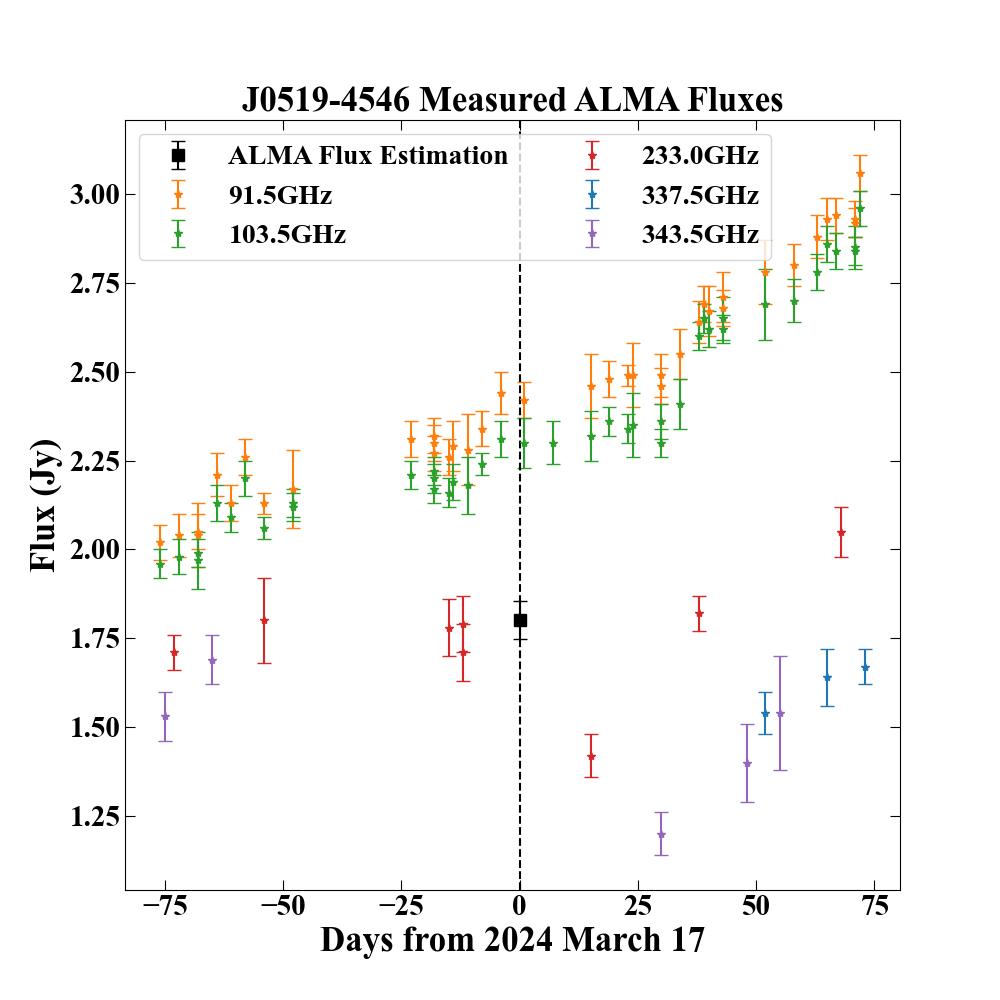}{0.50\textwidth}{(C)}
	}
\caption{\textbf{(A)--(C).} Measured and estimated fluxes for the quasar flux calibrators in this study, illustrating our estimation of the 233 GHz flux for J0538-4405 on March 8.
\label{fig:fluxcal}}
\end{figure*}

We compared the 233 GHz flux predicted by the ALMA Flux Estimation algorithm for each quasar on the date of observations against the nearby measured values. Figure~\ref{fig:fluxcal} shows our results. For J0334-4008 and J0519-4546, the predicted values used by the pipeline are in good agreement with nearby Band 6 measurements. However, the value for J0538-4405 (2.15 Jy; Execution 2 on March 8) is clearly lower than the trend in nearby measurements. Furthermore, extrapolating the Band 6 flux across the $\pm$80 day window using the ALMA Flux Estimation spectral index ($\alpha$ = -0.659) underestimates the measured Band 6 fluxes across the entire window. Instead, we fit fifth-order polynomials to the measured Band 3 and Band 6 fluxes, then used these to predict the Band 6 flux on March 8. Our procedure provides a flux of 2.38 Jy, which better fits the trend. We applied a correction to our measurement set to adjust the J0538-4405 flux scale to 2.38 Jy and correspondingly scaled the comet and phase calibrator fluxes (a factor of $\sim1.05$). 

%% file: AppendixObserving.tex
\section{Observing Log}\label{sec:obslog}

Table~\ref{tab:obslog} summarizes observing circumstances for the ALMA observations, and Table~\ref{tab:lco} provides observing circumstances and photometry for LCO.

\begin{deluxetable*}{cccccccc}[h]
\tablenum{B1}
\tablecaption{Observing Log\label{tab:obslog}}
\tablewidth{0pt}
\tablehead{
\colhead{Execution} & \colhead{UT Time} & \colhead{\textit{T}\subs{int}} &
 \colhead{$N$\subs{ants}} & \colhead{Baselines} & \colhead{PWV} & \colhead{$\theta$\subs{INT}} & \colhead{$\theta$\subs{INT}}  \\
\colhead{} & \colhead{} & \colhead{(min)} & \colhead{} & \colhead{(m)}  & \colhead{(mm)} & \colhead{($\arcsec$)} & \colhead{(km)}
}
\startdata
\multicolumn{8}{c}{UT 2024 March 8--9, \rh{}=16.61 au, $\Delta$=16.91 au, $\phi$\subs{STO}=$3.2\degr$, $\psi_{\sun} = 119.5\degr$} \\
1 & 20:46--21:43 & 49 & 44 & 15--312 & 1.61 &  1.61$\times$1.48 & 19,745$\times$18,151 \\ 
2 & 21:54--22:51 & 49 & 44 & 15--312 & 1.38 & 1.70$\times$1.51 & 20,849$\times$18,519 \\
3 & 23:21--00:18 & 49 & 48 & 15--312 & 1.23 & 2.28$\times$1.42 & 27,962$\times$17,415 \\
\hline
\multicolumn{8}{c}{UT 2024 March 17--18, \rh{}=16.58 au, $\Delta$=16.87 au, $\phi$\subs{STO}=$3.2\degr$, $\psi_{\sun} = 128.2\degr$} \\ 
5 &  19:34--20:31 & 49 & 43 & 15--314 & 2.15 & 1.61$\times$1.48 & 19,698$\times$18,108 \\
6 & 20:52--21:49 & 49 & 43 & 15--284 & 2.20 & 1.79$\times$1.56 & 21,901$\times$19,087 \\
7 & 23:26--00:23 & 49 & 43 & 15--284 & 1.81 & 2.55$\times$1.46 & 31,200$\times$17,863 
\enddata
\tablecomments{\textit{T}\subs{int} is the total on-source integration time. \textit{r}\subs{H}, $\Delta$, and $\phi_\mathrm{{STO}}$, and $\psi_{\sun}$ are the heliocentric distance, geocentric distance,
solar phase angle (Sun--Comet--Earth), and position angle of the projected Sun--comet vector in the plane of the sky, respectively, of UN271 at the time of observations.  \textit{N}\subs{ants}
is the number of antennas utilized during each observation, with the range of baseline lengths indicated for each. PWV is the mean precipitable water vapor at zenith during the observations. $\theta$\subs{min} is the angular resolution (synthesized beam) at $\nu$ given in arcseconds and in projected distance (km) at the geocentric distance of UN271. All observations were conducted with a correlator setup centered at $\nu$ = 230 GHz, where the extent of the ALMA primary beam is 25.3$\arcsec$ ($\sim$310,000 km projected distance at UN271).}
\end{deluxetable*}

\begin{deluxetable*}{cccccccccc}
\tablenum{B2}
    \tablecaption{Optical photometry of comet C/2014 UN271 (Bernardinelli-Bernstein)\label{tab:lco}}
    \tablehead{
        \colhead{Date}
        & \colhead{Site}
        & \colhead{Filter}
        & \colhead{$r_\mathrm{H}$}
        & \colhead{$\Delta$}
        & \colhead{$\phi_\mathrm{STO}$}
        & \colhead{Airmass}
        & \colhead{IQ}
        & \colhead{$m$}
        & \colhead{$\sigma$} \\
        \colhead{(UTC))}
        &
        &
        & \colhead{(au)}
        & \colhead{(au)}
        & \colhead{(\degr)}
        &
        &
        & \colhead{(mag)}
        & \colhead{(mag)}
    }
    \startdata
    2024 Feb 03 01:59 & Cerro Tololo & $g'$ & 16.734 & 16.983 & 3.24 & 1.4 & 2.1 & 17.800 & 0.063 \\
    2024 Feb 03 02:03 & Cerro Tololo & $r'$ & 16.734 & 16.983 & 3.24 & 1.4 & 1.9 & 17.348 & 0.056 \\
    2024 Feb 06 02:14 & Cerro Tololo & $g'$ & 16.723 & 16.981 & 3.23 & 1.5 & 2.5 & 17.788 & 0.066 \\
    2024 Feb 06 02:17 & Cerro Tololo & $r'$ & 16.723 & 16.981 & 3.23 & 1.5 & 2.3 & 17.338 & 0.059 \\
    2024 Feb 08 20:09 & South Africa & $g'$ & 16.713 & 16.979 & 3.23 & 1.5 & 2.3 & 17.823 & 0.064 \\
    2024 Feb 08 20:12 & South Africa & $r'$ & 16.713 & 16.979 & 3.23 & 1.5 & 2.2 & 17.319 & 0.063 \\
    2024 Feb 17 01:40 & Cerro Tololo & $g'$ & 16.685 & 16.968 & 3.22 & 1.5 & 2.2 & 17.887 & 0.070 \\
    2024 Feb 17 01:44 & Cerro Tololo & $r'$ & 16.685 & 16.968 & 3.22 & 1.5 & 2.0 & 17.435 & 0.063 \\
    2024 Feb 20 01:05 & Cerro Tololo & $g'$ & 16.674 & 16.962 & 3.22 & 1.4 & 2.1 & 17.857 & 0.066 \\
    2024 Feb 20 01:09 & Cerro Tololo & $r'$ & 16.674 & 16.962 & 3.22 & 1.5 & 2.0 & 17.386 & 0.066 \\
    2024 Feb 24 01:07 & Cerro Tololo & $g'$ & 16.660 & 16.954 & 3.22 & 1.5 & 3.2 & 17.942 & 0.074 \\
    2024 Feb 24 01:11 & Cerro Tololo & $r'$ & 16.660 & 16.954 & 3.22 & 1.5 & 4.0 & 17.517 & 0.072 \\
    2024 Feb 29 00:27 & Cerro Tololo & $g'$ & 16.643 & 16.942 & 3.22 & 1.4 & 5.4 & 17.374 & 0.059 \\
    2024 Feb 29 00:31 & Cerro Tololo & $r'$ & 16.643 & 16.942 & 3.22 & 1.4 & 4.5 & 16.945 & 0.058 \\
    2024 Mar 02 00:34 & Cerro Tololo & $g'$ & 16.636 & 16.936 & 3.22 & 1.5 & 4.0 & 17.387 & 0.059 \\
    2024 Mar 02 00:38 & Cerro Tololo & $r'$ & 16.636 & 16.936 & 3.22 & 1.5 & 3.8 & 16.941 & 0.059 \\
    2024 Mar 03 00:47 & Cerro Tololo & $g'$ & 16.632 & 16.933 & 3.22 & 1.5 & 2.3 & 17.374 & 0.055 \\
    2024 Mar 03 00:51 & Cerro Tololo & $r'$ & 16.632 & 16.933 & 3.22 & 1.5 & 2.1 & 16.882 & 0.061 \\
    2024 Mar 04 00:12 & Cerro Tololo & $g'$ & 16.629 & 16.930 & 3.23 & 1.4 & 4.5 & 17.388 & 0.061 \\
    2024 Mar 04 00:15 & Cerro Tololo & $r'$ & 16.629 & 16.930 & 3.23 & 1.4 & 4.3 & 16.953 & 0.057 \\
    2024 Mar 05 00:38 & Cerro Tololo & $g'$ & 16.625 & 16.927 & 3.23 & 1.5 & 3.2 & 17.435 & 0.059 \\
    2024 Mar 05 00:42 & Cerro Tololo & $r'$ & 16.625 & 16.927 & 3.23 & 1.5 & 2.9 & 16.989 & 0.056 \\
    2024 Mar 06 00:34 & Cerro Tololo & $g'$ & 16.622 & 16.924 & 3.23 & 1.5 & 2.5 & 17.444 & 0.060 \\
    2024 Mar 06 00:37 & Cerro Tololo & $r'$ & 16.622 & 16.924 & 3.23 & 1.5 & 2.4 & 16.983 & 0.053 \\
    2024 Mar 07 00:40 & Cerro Tololo & $g'$ & 16.618 & 16.921 & 3.23 & 1.5 & 2.5 & 17.435 & 0.058 \\
    2024 Mar 07 00:44 & Cerro Tololo & $r'$ & 16.618 & 16.921 & 3.23 & 1.6 & 2.3 & 16.986 & 0.054 \\
    2024 Mar 08 00:07 & Cerro Tololo & $g'$ & 16.615 & 16.917 & 3.23 & 1.4 & 2.5 & 17.448 & 0.055 \\
    2024 Mar 08 00:10 & Cerro Tololo & $r'$ & 16.615 & 16.917 & 3.23 & 1.5 & 2.4 & 17.000 & 0.051 \\
    2024 Mar 09 00:31 & Cerro Tololo & $g'$ & 16.611 & 16.914 & 3.23 & 1.5 & 2.1 & 17.394 & 0.055 \\
    2024 Mar 09 00:35 & Cerro Tololo & $r'$ & 16.611 & 16.914 & 3.23 & 1.6 & 2.0 & 16.961 & 0.049 \\
    2024 Mar 10 00:38 & Cerro Tololo & $g'$ & 16.608 & 16.910 & 3.23 & 1.6 & 2.7 & 17.461 & 0.057 \\
    2024 Mar 10 00:42 & Cerro Tololo & $r'$ & 16.608 & 16.910 & 3.23 & 1.6 & 2.6 & 17.028 & 0.050 \\
    2024 Mar 18 17:58 & South Africa & $g'$ & 16.577 & 16.876 & 3.25 & 1.5 & 2.2 & 17.633 & 0.060 \\
    2024 Mar 18 18:01 & South Africa & $r'$ & 16.577 & 16.876 & 3.25 & 1.5 & 2.1 & 17.168 & 0.055 \\
    2024 Mar 19 10:51 & Siding Spring & $g'$ & 16.575 & 16.873 & 3.25 & 2.0 & 2.3 & 17.701 & 0.082 \\
    2024 Mar 19 10:51 & Siding Spring & $r'$ & 16.575 & 16.873 & 3.25 & 2.0 & 2.0 & 17.231 & 0.069 \\
    2024 Mar 20 00:01 & Cerro Tololo & $g'$ & 16.573 & 16.871 & 3.25 & 1.6 & 3.4 & 17.744 & 0.063 \\
    2024 Mar 20 00:05 & Cerro Tololo & $r'$ & 16.573 & 16.871 & 3.25 & 1.6 & 3.0 & 17.261 & 0.057 \\
    2024 Mar 23 23:56 & Cerro Tololo & $g'$ & 16.559 & 16.852 & 3.27 & 1.6 & 2.6 & 17.565 & 0.061 \\
    2024 Mar 24 00:00 & Cerro Tololo & $r'$ & 16.559 & 16.852 & 3.27 & 1.6 & 2.6 & 17.143 & 0.058
    \enddata
    \tablecomments{Columns: (Filter) filter used for the observation; (IQ) image
        quality, i.e., FWHM of point sources; ($m$) apparent magnitude in 6\arcsec{}
        radius, calibrated to the PS1 photometric system, $g$ for the $g'$ filter
        and $r$ for the $r'$ filter' ($\sigma$) 1$\sigma$ uncertainty on $m$.}
\end{deluxetable*}

%% file: AppendixGas.tex
\section{CO Fourier Domain Modeling}\label{sec:fourier}
We chose to perform modeling of the interferometric spectra in the Fourier domain following \cite{Cordiner2023} to avoid the introduction of imaging artifacts owing to the incomplete $uv$ sampling inherent to interferometric observations. We used the \texttt{vis\_sample} program \citep{Loomis2018} to take the Fourier transform of our radiative transfer model cubes using the same $uv$ coverage as our ALMA observations. We then performed least-squares fits of our radiative transfer models against the measured interferometric visibilities. We used the \texttt{lmfit} application of the Levenberg-Marquardt minimization technique and retrieved uncertainties on our optimized parameters from the diagonal elements of the covariance matrix, minimizing the residual between both the real and imaginary parts of the observed and modeled interferometric visibilities. 

\begin{figure*}
\gridline{\fig{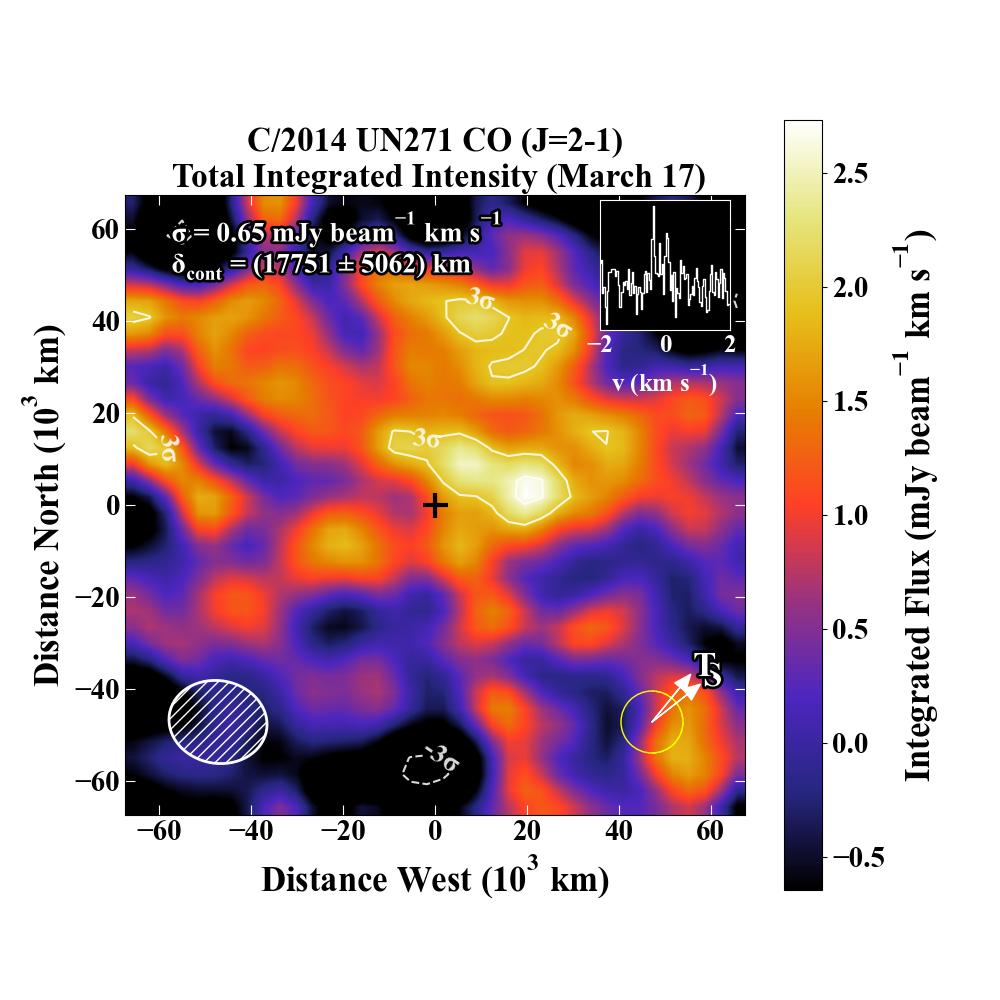}{0.45\textwidth}{(A)}
          \fig{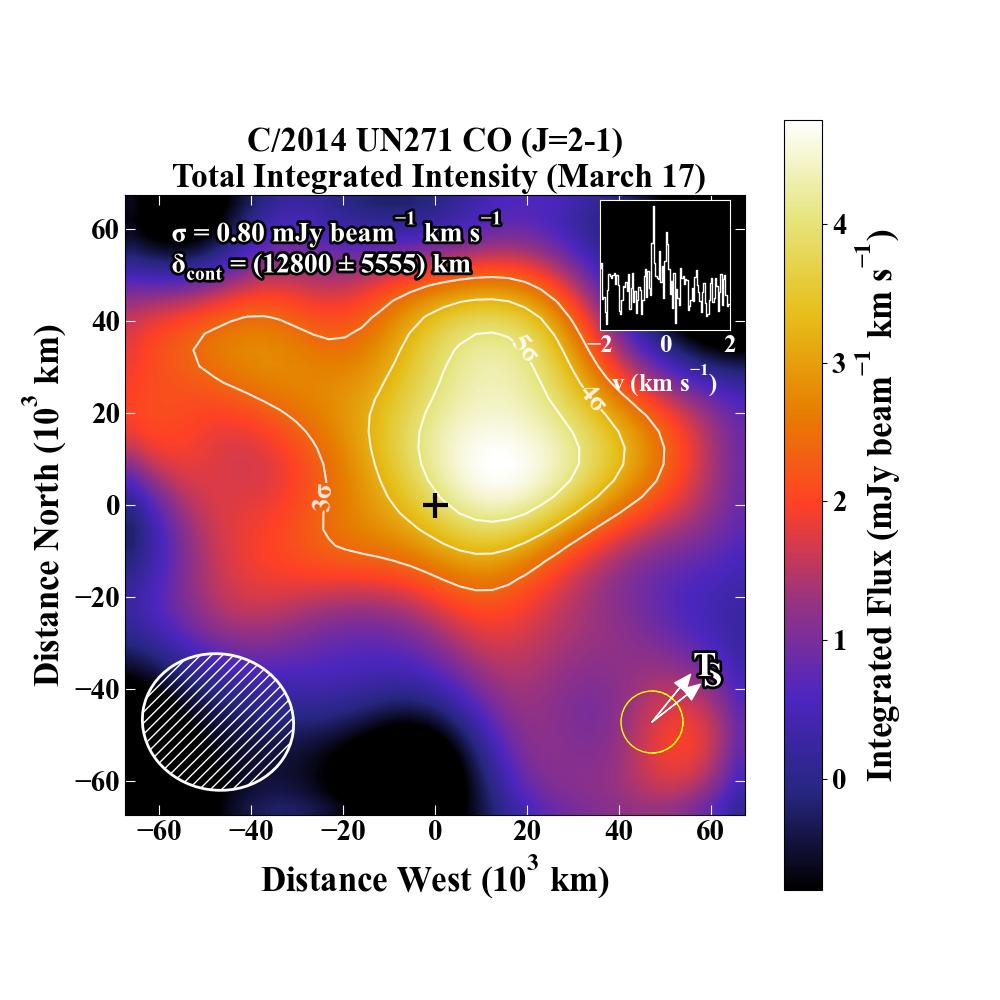}{0.45\textwidth}{(B)}
}
\gridline{\fig{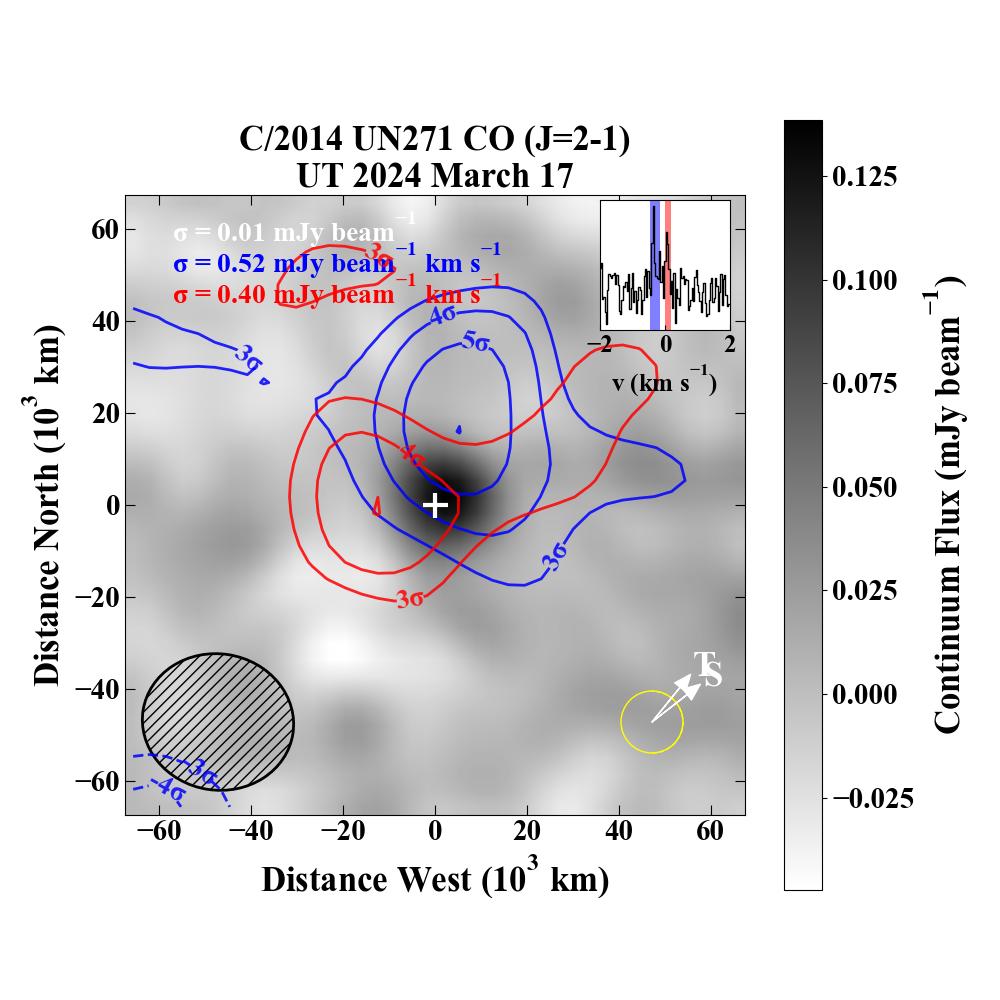}{0.45\textwidth}{(C)}
          \fig{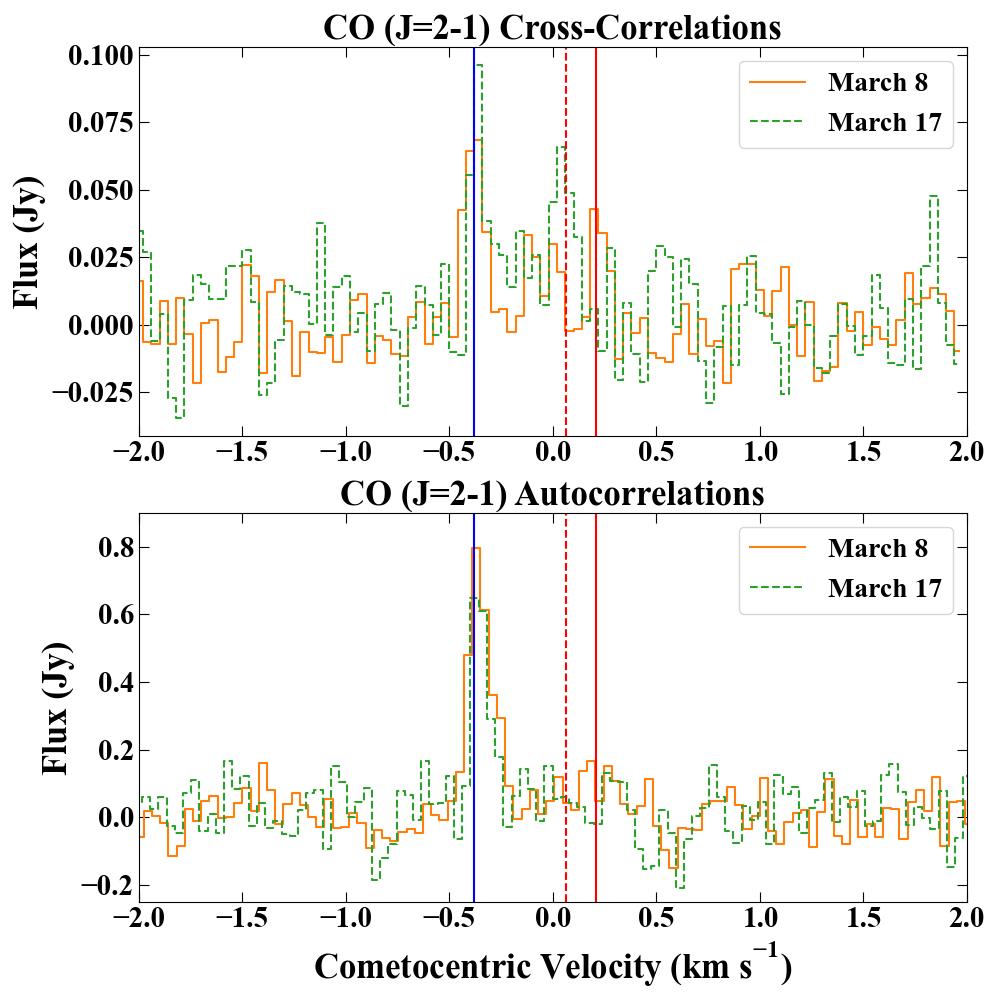}{0.45\textwidth}{(D)}
}
\caption{\textbf{(A)--(C).} Spectrally integrated flux maps for CO on 2024 March 17, with traces and labels as in Figure~\ref{fig:maps}. Contour intervals in each map are in 1$\sigma$ increments of the rms noise, with the lowest contour being 3$\sigma$. Panel (A) is at the native angular resolution, whereas Panel (B) shows the same map with a 2$\arcsec$ Gaussian $uv$-taper. \textbf{(D).} Comparison of autocorrelation and interferometric CO spectra on March 8 and 17. The solid blue, dashed red, and solid red vertical lines show correspond to speeds of -0.38 \kms{}, 0.066 \kms{}, and 0.21 \kms{}, respectively, highlighting the consistency of the blue CO component and variation in the red components.
\label{fig:maps17}}
\end{figure*}

Following the formalism described in \S~\ref{subsec:kinematics}, we divided the coma into two regions, $R_1$ and $R_2$. Region $R_1$ is defined as the conical region with half-opening angle $\gamma$, originating at the nucleus surface and oriented at a phase angle $\phi$ with respect to the observer and position angle $\psi$ in the plane of the sky \citep[see Fig. 8 of][]{Cordiner2023}. We assumed a gas kinetic temperature $T_{\mathrm{kin}}$ = 4 K given the large \rh{} of UN271 and measures in other distant comets \citep[\eg{}][]{Gunnarsson2008,Roth2023}. We determined the CO distribution in $R_1, R_2$ using a Haser formalism \citep{Haser1957}:
\begin{equation}
    n_d(r) = \frac{Q_i}{4 \pi v_i r^2}\frac{\frac{v_i}{\beta_d}}{\frac{v_i}{\beta_p}-\frac{v_i}{\beta_d}}\left[\exp{\left(-\frac{\beta_p}{v_i}r\right)}-\exp{\left(-\frac{\beta_d}{v_i}r\right)}\right],
\end{equation}
\noindent where $Q_i$ and $v_i$ are the molecular production rate (s$^{-1}$) and gas expansion velocity (km\,s$^{-1}$) in each coma region, and $\beta_p, \beta_d$ are the parent and daughter molecular photodissociation rates (s$^{-1}$), respectively. 

We first worked to determine the outgassing geometry, assuming direct nucleus release (\textit{L}\subs{p} = 0 km) for CO and allowing the molecular production rate, expansion speed, and jet geometry ($Q_1, Q_2, v_1, v_2, \gamma, \phi, \psi$) to vary as free parameters. The integrated flux maps (Fig.~\ref{fig:maps}) provided strong constraints on the geometry, including a blueward jet roughly along the projected anti-sunward direction (where $\psi_{\sun} = 119\degr$). The high spectral resolution  and sensitivity provided by ALMA across the narrow CO line provided tight constraints on our model, with the jet best described by a half-opening angle $\gamma = (27 \pm 5)\degr$, $\phi = (2.1 \pm 1.0)\degr$, $\psi = (306 \pm 10)\degr$, and a ratio of production rates in the jet vs.\ ambient coma $Q_1/Q_2 = 6.2\pm1.4$ on March 8. On March 17, a lower signal-to-noise ratio prevented firm constraints on the phase and position angles of $R_1$ (Figure~\ref{fig:maps17}). We assumed that it was oriented along the Sun-comet line. Our results depict a very different outgassing geometry, with the half-opening angle $\gamma = (73 \pm 7)\degr$ and $Q_1/Q_2 = (28\pm 13)$.

\begin{figure*}
    \gridline{\fig{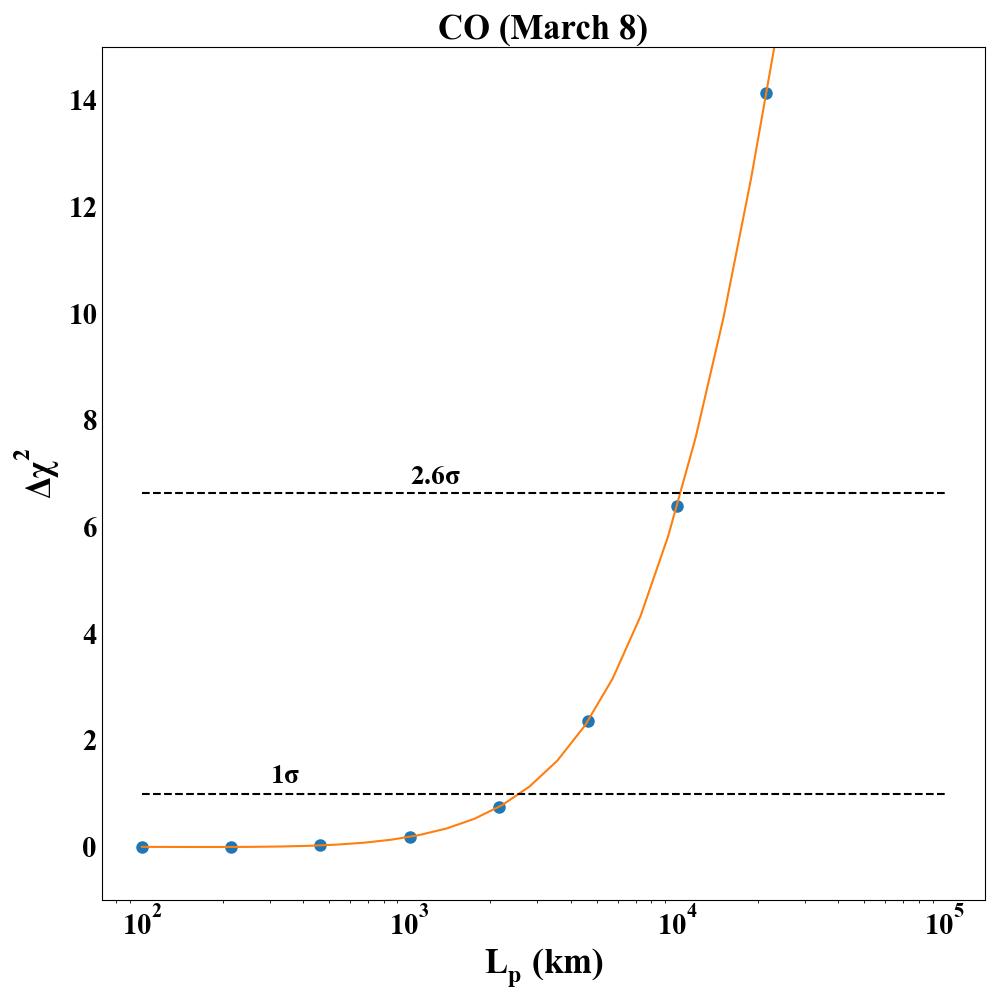}{0.35\textwidth}{(A)}
              \fig{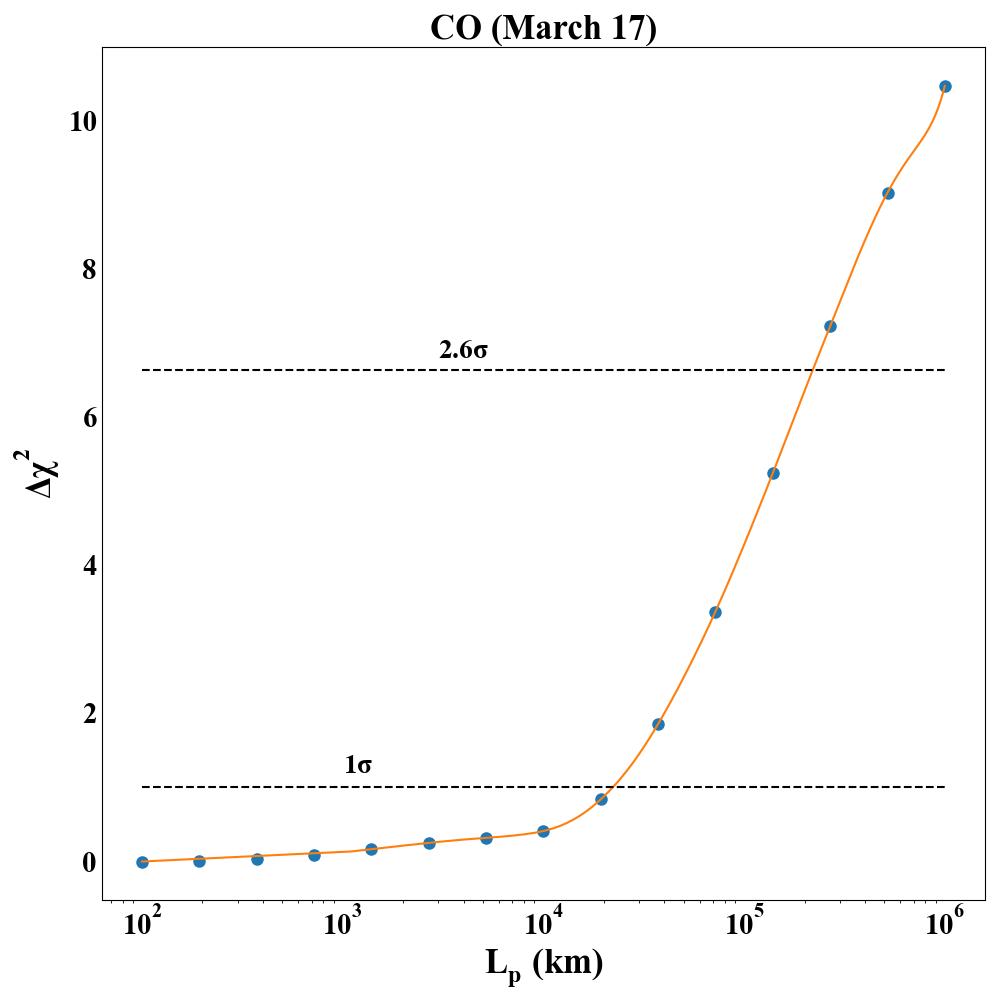}{0.35\textwidth}{(B)}
    }
\caption{\textbf{(A--B).} $\Delta\chi^2$ analysis for the CO $L_p$ in UN271 on March 8 and 17.
\label{fig:dchisq}}
\end{figure*}

We then worked to calculate the parent scale length for CO to test for contributions from direct nucleus release vs.\ coma chemistry, including CO$_2$ or H$_2$CO photolysis. The density of coma molecules peaks at the nucleus before falling off exponentially with increasing nucleocentric distance owing to adiabatic expansion and photolysis. To adequately sample the often asymmetric uncertainties on $L_p$, we generated the $\chi^2$ surface for CO as a function of $L_p$, allowing the production rate and expansion speed to vary but holding all other parameters fixed for each value of $L_p$. We then plotted the $\Delta\chi^2$ curve generated using cubic spline interpolation between each value of $L_p$, where $\Delta\chi^2(L_p) = \chi^2(L_p) - \chi^2_{min}$. We obtained the 1$\sigma$ and 2.6$\sigma$ uncertainties from the values for $\Delta\chi^2$ = 1 (68\% confidence) and 6.63 (99\% confidence) thresholds. Figure~\ref{fig:dchisq} shows our $\Delta\chi^2$ analysis for each epoch.

%% file: AppendixActive.tex
\section{Active CO Fraction}\label{sec:active}
We calculated the CO active fractional area for UN271 and Hale-Bopp using the sublimation model of \cite{Cowan1979} for a visual abledo of 0.05 and an infrared emissivity of 0.95. We used the Small-Bodies-Node/ice-sublimation code \citep{VanSelous2021} to calculate the average CO sublimation rate per surface unit, $Z$, at \rh{} = 16.6 au for UN271 and \rh{} = 14 au for Hale-Bopp. We used $Q$(CO) for UN271 reported here and $Q$(CO) = $3\times10^{27}$ s$^{-1}$ for Hale-Bopp at \rh{} = 14 au post-perihleion \citep{Biver2002}. The active area is calculated by dividing $Q$(CO) by $Z$, and the nucleus active fractional area is found by dividing the active area by the nucleus surface area. Our results are given in Table~\ref{tab:active}.

\begin{deluxetable*}{cccccc}
\tablenum{D1}
\tablecaption{C/2014 UN271 and C/1995 O1 (Hale-Bopp) CO Active Areas \label{tab:active}}
\tablewidth{0pt}
\tablehead{
\colhead{Comet} & \colhead{$Q$(CO)\sups{a}} & \colhead{$Z$} & \colhead{$d_N$\sups{b}} & \colhead{$A$} & \colhead{$f$} \\
\colhead{} & \colhead{(mol s$^{-1}$)} & \colhead{(mol m$^{-2}$ s$^{-1}$)} & \colhead{(km)} & \colhead{(km$^2$)} & \colhead{(\%)}
}
\startdata
\hline
UN271 (\rh{}=16.6 au) & $3.44\times10^{27}$ & $8.4\times10^{19}$ & 137 & 41 & 0.07 \\
Hale-Bopp (\rh{}=14 au) & $3\times10^{27}$ & $1.2\times10^{20}$ & 60 & 25 & 0.22 
\enddata  
\tablecomments{\sups{a}$Q$(CO) as reported for UN271 in this work and in \cite{Biver2002} for Hale-Bopp. \sups{b} Nucleus diameter reported for UN271 in \cite{Lellouch2022} and an intermediate value of the range reported for Hale-Bopp in \cite{Weaver1997}.
}
\end{deluxetable*}

%% file: AppendixContinuum.tex
\section{Continuum Modeling}\label{sec:cont_analysis}
We chose to model the continuum emission in the Fourier domain for similar reasons as explained for the CO gas (Appendix~\ref{sec:fourier}) following the methods of \cite{Lellouch2022}, including the use of the Near Earth Asteroid Thermal Model \citep[NEATM;][]{Harris1998,Delbo2002}. Here we recount the formalism of these manuscripts for completeness and describe their application to our study of UN271.

\begin{figure*}
\gridline{\fig{UN271.cont.Mar8.233GHz.jpg}{0.45\textwidth}{(A)}
          \fig{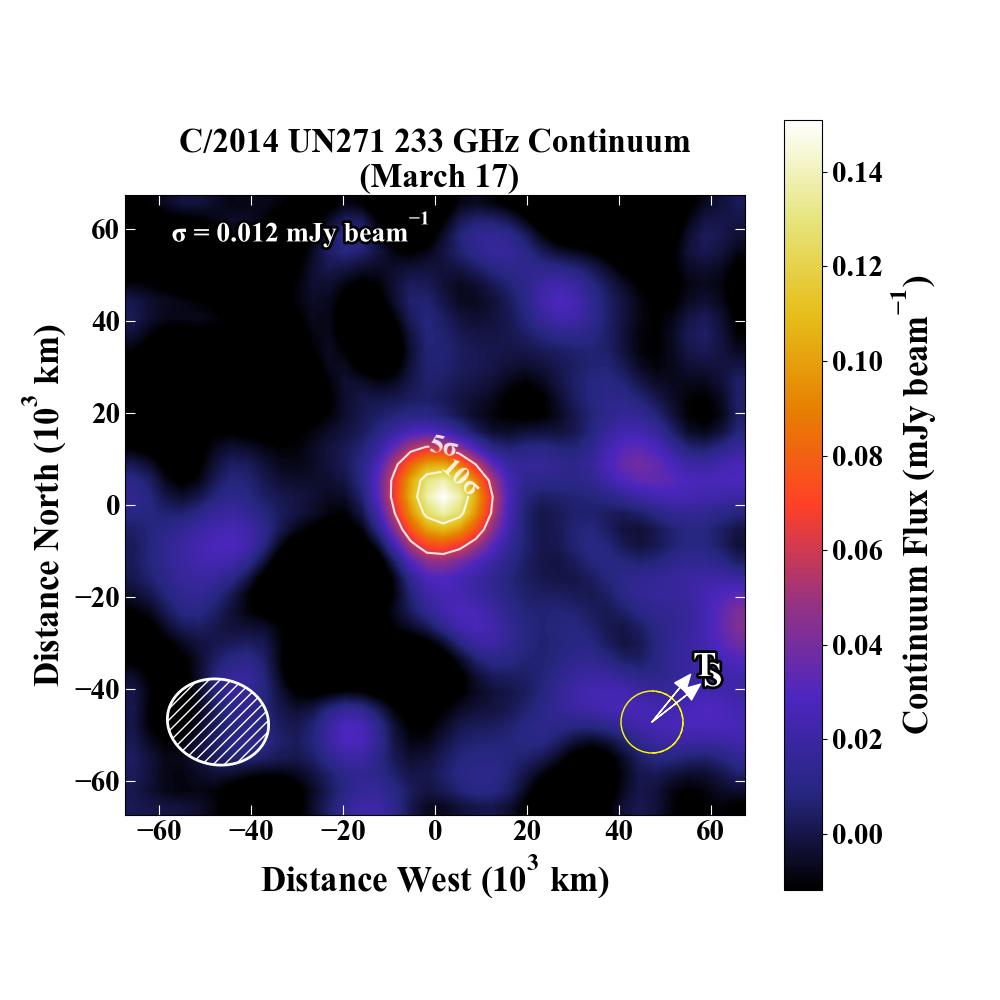}{0.45\textwidth}{(B)}
}
\caption{\textbf{(A)--(B).} Continuum flux maps on 2024 March 8 and 17, with traces and labels as in Figure~\ref{fig:maps}). Contour intervals in each map are in 5$\sigma$ increments of the rms noise, with the lowest contour being 5$\sigma$. 
\label{fig:cont_compare}}
\end{figure*}

The continuum images and visibilities revealed no signs of extended emission (Figures~\ref{fig:cont_fits} and~\ref{fig:cont_compare}). In the Fourier domain, an unresolved point source ($\delta$ function) would show a constant value with $uv$-radius ($\sigma$), whereas the visibility amplitude $V$ for a coma following a $\rho^{-1}$ distribution should vary as 

\begin{equation}\label{eq:eq1}
V(\sigma)=\frac{Kc}{\nu\sigma}
\end{equation} 

\noindent where K is a constant accounting for the emission properties of the dust and

\begin{equation}\label{eq:eq2}
V(0) = K\sqrt{\frac{\pi}{4\ln 2}}\pi\Phi
\end{equation}

\noindent where $\Phi$ is the primary beam width in radians. K can be derived by fits to the visibilities and in turn related to the dust mass through the total measured flux (Equation~\ref{eq:eq5}). 

We used the \texttt{lmfit} package to perform least squares fits to the 233 GHz continuum visibilities using (1) A point source model, (2) A point source model and a coma model, and (3) A point source model with a fixed coma model derived from the 3$\sigma$ upper limit on the shortest baseline flux. Our formalism for interpreting the best-fit components of each model is given below.

\subsection{Nucleus Size}\label{subsec:nucleus}
The NEATM \citep{Harris1998,Delbo2002} was used to calculate the nucleus diameter implied by a given point-source component for each model fit. The NEATM accounts for the solar phase angle ($\alpha$) during observations by considering the observed flux originating from the illuminated portion of the nucleus from the observer's perspective. We calculated the temperature at the sub-solar point ($T_{ss}$) as

\begin{equation}\label{eq:eq2}
T_{ss} = \left[\frac{(1-A)S_{\sun}}{\rh{}^2\epsilon\sigma\eta}\right]^{1/4}
\end{equation}

\noindent where A is the Bond albedo, S$_{\sun}$ is the Solar constant at 1 au (1360.8 W\,m$^{-2}$), \rh{} is the heliocentric distance (au), $\sigma$ is the Stefan Boltzmann constant $(5.67\times10^{-8}$ J\,s$^{-1}$\,m$^{-2}$\,K$^{-4}$), $\epsilon$ is the bolometric emissivity, and $\eta$ the infrared beaming factor. The resulting $T_{ss}$ is 95 $\pm$ 9 K on both dates. This is higher than the $T_{ss}$ calculated from the CO expansion speed (60 K; Section~\ref{subsubsec:interpretation}). Therefore, we performed calculations in both cases, first using $T_{ss}$ derived from the nucleus parameters of \cite{Lellouch2022} and next using $T_{ss}$ as calculated from our CO spectra. The temperature was then calculated across the nucleus surface assuming that no emission originates from the night side as

\begin{equation}\label{eq:eq3}
T(\theta,\phi) = T_{ss}(\cos\phi\cos\theta)^{1/4}
\end{equation}

\noindent for $\theta \in [\alpha-\pi/2,\alpha+\pi/2]$ and $T(\theta,\phi) = 0$ elsewhere. The expected flux ($F_{\lambda}$) at a wavelength $\lambda$ was then calculated as

\begin{equation}\label{eq:eq4}
F_\lambda = \frac{\epsilon(\lambda)d^2}{\Delta^2}\frac{hc^2}{\lambda^5} \int_0^{\pi/2} \int_{-\pi/2}^{\pi/2} \frac{\cos^2\phi\cos(\theta-\alpha)}{\exp\left( \frac{hc}{\lambda k_BT(\theta,\phi)} \right)-1} \mathrm{d}\theta\mathrm{d}\phi
\end{equation}

\noindent where $F_{\lambda}$ is the measured flux density (with units of W m$^{-2}$ $\mu$m$^{-1}$), $\epsilon(\lambda)$ is the wavelength-dependent spectral emissivity, $k_B$ is the Boltzmann constant ($1.38\times10^{-23}$ J\,K$^{-1}$), $d$ is the comet nucleus diameter (m), and $\Delta$ the geocentric distance (m). 

\subsection{Coma Dust Mass}\label{subsec:coma_limits}
For deriving the coma dust mass we followed previous methods \citep{Lellouch2022,Roth2021a,Boissier2012}, where the dust mass (M, kg) is related to the measured flux (S$_\lambda$, Jy) as

\begin{equation}\label{eq:eq5}
S_\lambda = \frac{2k_BTM\kappa_\lambda}{\lambda^2\Delta^2}
\end{equation}

where T is the dust temperature (taken to be the equilibrium temperature for grains at \rh{} = 16.6 au) and $\kappa_\lambda$ (m$^2$ kg$^{-1}$) is the dust opacity, which expresses the effective surface area for absorption available per unit mass and is related to the optical depth of the material. We adopted dust opacities calculated by \cite{Boissier2012} at $\lambda$ = 1.2 mm for a size index of -3, a maximum size of 1 mm, and a dust porosity of 50\%. \cite{Lellouch2022} calculated a maximum liftable particle size of 8 $\mu$m from the surface of UN271 for $Q$(CO) = $7\times10^{27}$ s$^{-1}$, nearly twice our measured value on March 8; thus, we consider the maximum particle size of 1 mm to provide a conservative estimate for the dust flux during quiescent activity. However, this maximum liftable size does not necessarily apply during outbursts, and large particles were a significant component of the 17P/Holmes outburst dust cloud \citep{Ishiguro2016}. For optical constants corresponding to astronomical (amorphous) silicates or silicate mixtures made of 20\% amorphous silicates and 80\% forsterite, mixed with water ice in various proportions, the derived opacities are in the range $1.70-4.67\times10^{-1}$ m$^2$ kg$^{-1}$. Dust masses are provided for the upper and lower range of these opacities. Table~\ref{tab:cont} provides our best-fit nucleus sizes and dust coma masses for each model scenario.

\begin{deluxetable*}{cccccc}
\tablenum{E1}
\tablecaption{Continuum Emission Analysis in C/2014 UN271 \label{tab:cont}}
\tablewidth{0pt}
\tablehead{
\colhead{Model} & \colhead{$F_n$\sups{a}} & \colhead{$T_{\mathrm{ss}}$\sups{b}} & \colhead{$D_n$\sups{c}} & \colhead{$K$\sups{d}} & \colhead{$M_d$\sups{e}} \\
\colhead{} & \colhead{(mJy)} & \colhead{(K)} & \colhead{(km)} & \colhead{(Jy)} & \colhead{($10^{11}$ kg)}
}
\startdata
\multicolumn{6}{c}{2024 March 8, $T_{\mathrm{ss}}$(NEATM) = $95\pm9$ K, $T_{\mathrm{ss}}$(CO) = $60\pm2$ K} \\
\hline
Point Source & 95 & 0.130 $\pm$ 0.013 & 107 $\pm$ 11 & ... & ... \\
             & 60 & (0.130)           & 129 $\pm$ 14 & ... & ... \\
Point$+$Coma & 95 & 0.114 $\pm$ 0.012 & 101 $\pm$ 11 & 0.88 $\pm$ 0.16 & $(0.44\pm0.08$)--$(1.2\pm0.2)$ \\
             & 60 & (0.114) & 129 $\pm$ 14 & ... & ... \\
Max Coma & ... & ... & ... & $<$1.9 $(3\sigma)$ & $<$0.9--2.5 $(3\sigma)$ \\
\hline
\multicolumn{6}{c}{2024 March 17, $T_{\mathrm{ss}}$(NEATM) = $95\pm9$ K, $T_{\mathrm{ss}}$(CO) = $60\pm3$ K} \\
\hline
Point Source & 95 & 0.157 $\pm$ 0.016 & 118 $\pm$ 12 & ... & ... \\
             & 60 & (0.157)           & 151 $\pm$ 16 & ... & ... \\
Point$+$Coma & 95 & 0.157 $\pm$ 0.016 & 118 $\pm$ 13 & $<$0.3 ($3\sigma$) & $<$0.1--0.4 (3$\sigma$) \\
             & 60 & (0.157)            & 151 $\pm$ 16 & ... & ... \\
Max Coma     & ... & ... & ... & $<$0.5 $(3\sigma)$ & $<$0.2--0.5 $(3\sigma)$
\enddata  
\tablecomments{\sups{a} Best-fit point source flux. \sups{b} Subsolar temperature used to calculate nucleus size. \sups{c} Calculated nucleus diameter. \sups{d} Coma model visibility coefficient (Appendix~\ref{sec:cont_analysis}). \sups{e} Calculated dust masses using a range of dust opacities (Appendix~\ref{sec:cont_analysis}). 
}
\end{deluxetable*}